\documentclass{aa}  

\usepackage{graphicx}
\usepackage{txfonts}
\begin{document}

\title{Milky Way demographics with the VVV survey. }

\subtitle{IV. PSF photometry from almost one billion stars in the Galactic bulge and adjacent southern disk.}

\author{Javier Alonso-Garc\'{i}a
          \inst{1,}\inst{2}
          \and
          Roberto K. Saito
          \inst{3}
          \and
          Maren Hempel
          \inst{4}
          \and
          Dante Minniti
          \inst{5,}\inst{2,}\inst{6}
          \and
          Joyce Pullen
          \inst{2}
          \and
          M\'{a}rcio Catelan
          \inst{4,}\inst{2,}\thanks{On sabbatical leave at European Southern Observatory, Alonso de C\'ordova 3107, Vitacura, Santiago, Chile}
          \and
          Rodrigo Contreras Ramos
          \inst{2,}\inst{4}
          \and
          Nicholas J. G. Cross
          \inst{7}
          \and
          Oscar A. Gonzalez
          \inst{8}
          \and
          Philip W. Lucas
          \inst{9}
          \and
          Tali Palma
          \and
          \inst{10}
          Elena Valenti
          \inst{11}
          \and
          Manuela Zoccali
          \inst{4,}\inst{2}
          }

    \institute{Centro de Astronom\'{i}a (CITEVA), Universidad de Antofagasta, Av. Angamos 601, Antofagasta, Chile\\
       \email{javier.alonso@uantof.cl}
         \and
         Instituto Milenio de Astrof\'{i}sica, Santiago, Chile
         \and
         Departamento de F\'{i}sica, Universidade Federal de Santa Catarina, Trindade 88040-900, Florian\'{o}polis, SC, Brazil
         \and
         Instituto de Astrof\'{i}sica, Pontificia Universidad Cat\'{o}lica de Chile, Av. Vicu\~{n}a Mackenna 4860, 782-0436 Macul, Santiago, Chile
         \and
         Departamento de F\'{i}sica, Facultad de Ciencias Exactas, Universidad Andr\'{e}s Bello, Av. Fern\'{a}ndez Concha 700, Las Condes, Santiago, Chile
         \and
         Vatican Observatory, V-00120 Vatican City State, Italy
         \and
         SUPA (Scottish Universities Physics Alliance) Wide-Field Astronomy Unit, Institute for Astronomy, School of Physics and Astronomy, University of Edinburgh, Royal Observatory, Blackford Hill, Edinburgh, EH9 3HJ, UK
         \and
         UK Astronomy Technology Centre, Royal Observatory, Blackford Hill, Edinburgh EH9 3HJ, UK
         \and
         Centre for Astrophysics, University of Hertfordshire, Hatfield AL10 9AB, UK
         \and
         Observatorio Astron\'{o}mico de C\'{o}rdoba, Universidad Nacional de C\'{o}rdoba, Laprida 854, X5000BGR, C\'{o}rdoba, Argentina
         \and
         European Southern Observatory, Karl-Schwarszchild-Str. 2, D-85748 Garching bei Muenchen, Germany
             }



\newpage
    
  \abstract
  {The inner regions of the Galaxy are severely affected by extinction, which limits our capability to study the stellar populations present there. The Vista Variables in the V\'{i}a L\'actea (VVV) ESO Public Survey has observed this zone at near-infrared wavelengths where reddening is highly diminished.}
  {By exploiting the high resolution and wide field-of-view of the VVV images we aim to produce a deep, homogeneous, and highly complete database of sources that cover the innermost regions of our Galaxy.}
  {To better deal with the high crowding in the surveyed areas, we have used point spread function (PSF)-fitting techniques to obtain a new photometry of the VVV images, in the $ZYJHK_s$ near-infrared filters available.}
   {Our final catalogs contain close to one billion sources, with precise photometry in up to five near-infrared filters, and they are already being used to provide an unprecedented view of the inner Galactic stellar populations. We make these catalogs publicly available to the community. 
     Our catalogs allow us to build the VVV giga-CMD, a series of color-magnitude diagrams of the inner regions of the Milky Way presented as supplementary videos.
     We provide a qualitative analysis of some representative CMDs of the inner regions of the Galaxy, and briefly mention some of the studies we have developed with this new dataset so far.}
   {}


   \maketitle
%

\section{Introduction}
\label{sec_intro}
In principle, the high stellar densities in the inner regions of our Galaxy and their relative closeness should produce the kind of well populated color-magnitude diagrams (CMDs) that are ideal to study their stellar populations. In practice, however, stars located at low Galactic latitudes in the inner parts of the Milky Way are hidden behind a curtain of dust and gas that highly extinguish their emission at optical and shorter wavelengths. Near-infrared observations are better suited for studies in these regions due to the diminished effect of extinction at these wavelengths ($A_{K_s}\sim 0.1A_V$). But until recent years, the kind of wide-field, near-infrared telescopes and cameras necessary to survey these relatively big regions of sky were not available. Large surveys in the near-infrared such as 2MASS and dedicated facilities like the VISTA telescope and its imager have completely changed this situation. Nowadays, the product of 4m aperture and 0.6 square degrees sky coverage per pointing makes VISTA the fastest near-infrared survey system in the world \citep{sut15}. The Vista Variables in the V\'ia L\'actea (VVV) survey, one of the six original key ESO public surveys conducted in Paranal with the VISTA telescope \citep{min10,sai12b}, takes full advantage of this fact to provide a new view of the inner regions of our Galaxy.

Our collaboration pioneered the use of the ESO public photometric catalogs available from VVV to provide a wide view of the stellar populations residing in the Galactic bulge region \citep{sai12a}, and in an adjacent Galactic disk region \citep{sot13}. However, current catalogs, based on aperture photometry, are unable to exploit the full potential of the VVV images, due to the high crowding present in the inner Galactic environments. Point spread function (PSF) photometry can provide a much more complete picture in the most crowded regions surveyed by VVV, such as the Galactic center or the inner Galactic globular clusters. But even the less crowded VVV regions can benefit from using PSF photometry by highly increasing the number of detected sources, as we show in this work.

\section {Observations}
\label{sec_obs}
The VVV observations were taken with the VIRCAM camera on the 4.1m VISTA telescope located in Cerro Paranal Observatory, in Chile. The VVV surveyed regions include the portion of sky located between $-10\fdg0 \le l \le +10\fdg4$ and $-10\fdg3 \le b \le +5\fdg1$ for the Galactic bulge, and between $294\fdg7 \le l\le350\fdg0$ and $-2\fdg25 \le b \le 2\fdg25$ for the low-latitude Galactic disk (see Figure \ref{fig_dens}). They were observed over a six-year period (2010-2015) with the $K_s$ filter, which was used for the variability campaign, between 69 and 293 times in the Galactic bulge area, and between 48 and 52 times in the disk area. All the VVV surveyed regions were also observed at least twice in the $Z$,$Y$,$J$, and $H$ filters, a first epoch in 2010-2011, and a second one in 2015. The VVV observations are divided in 196 contiguous fields in the Galactic bulge and 152 contiguous fields in an adjacent region in the southern disk. The VIRCAM camera contains 16 detectors, each one with $2048\times2048$ pixels. The pixel size is $\sim0.34''$. The detectors in the VIRCAM camera have significant gaps between them, generating so-called pawprint images. The observing strategy of the VVV survey, described in detail in \citet{sai12b}, consists in firstly taking a set of two slightly jittered images to account for detector cosmetic effects. This jittering is $\sim20\arcsec$ in both coordinates of the detector. The combination of these two images generates the so-called stacked pawprints. Additionally, in order to have a complete coverage of the area of every field, we take six consecutive stacked pawprints, dithered following a mosaic pattern to cover all the gaps. The combination of these stacked pawprints produces a full image of the field, a so-called tile. The area covered by a single tile is $1.5\times1.1$ square degrees in the sky, and each pixel in the tile, except for the borders, have been exposed at least four times.

The VVV observations are reduced, combined in stacked pawprints and tiles, astrometrized and calibrated by the Cambridge Astronomical Survey Unit (CASU; \citealt{eme04,irw04,ham04}). CASU also provides a catalog of aperture photometry, both for the stacked pawprint and for the tile images. However, as mentioned in Section \ref{sec_intro}, PSF photometry is better suited to obtain optimal results in the high-stellar-density regions VVV scanned \citep{alo15}. Still, as we show in Section \ref{sec_psf}, our PSF photometry makes use of the stacked science images produced by CASU, and the calibration of our PSF photometry relies heavily on the astrometric and photometric solutions provided by CASU. 

\section{PSF photometry and catalogs}
\label{sec_psf}
We proceeded by performing PSF photometry on the individual chips of the VVV stacked pawprints using DoPHOT \citep{sch93,alo12}. Effective exposure times of the analyzed VVV stacked images are detailed in Table \ref{tab_exptimes}. We used the stacked pawprints provided by CASU because they are less noisy than single pawprint images, and they do not show the abrupt and difficult-to-model PSF variations observed in tiles \citep{alo15}. Additionally, independent runs of DoPHOT in every chip of the stacked pawprints in the different fields allowed us to inject every time physical parameters that change with the night conditions, for example, full width at half maximum of the stellar sources, average sky counts; and others that depend on the individual detectors, for example, saturation limit. We flagged the borders from every chip of the stacked pawprints that were observed only once because of the jitter pattern (see Section \ref{sec_obs}). We avoided running DoPHOT on them to avert problems due to different input parameters and noise patterns with respect to the rest of the image that was effectively observed twice in the sequence. We transformed the instrumental positions of the sources reported by DoPHOT into ecliptic coordinates using WCSTools and the astrometric information provided by CASU for the stacked pawprints. This information allows for very small rms in the WCS ($\sim70$ mas, according to \citet{sai12b}). We also calibrated the instrumental photometry provided by DoPHOT by cross-matching it with the one provided in the CASU catalogs. A significant sample of the brightest, but non-saturated, sources in every chip (usually several thousands, always more than one hundred stars) was used to generate a zero-point offset that was applied to our PSF photometry to bring it to the VISTA photometric system \citep{gonfer18}. The process was parallelized in our local computing cluster to run it in over 300,000 images necessary to get the whole VVV region in the five near-infrared filters available, in two epochs per filter. We paid special attention to select the best quality VVV stacked images, whenever there were more than two epochs available for a given filter, and in particular for $K_s$ given the significant number of epochs available for this filter (see Section \ref{sec_obs}). To make the selection, we examined the seeing, ellipticity, and limit magnitude of the sources as provided by CASU and by our own pipeline. As we show in Figure \ref{fig_quality}, this resulted in using images with a most frequent seeing of $\sim0.75''$ and ellipticity of $\sim0.07$, with small differences in the magnitude limits between the two different epochs of less than $\sim0.2$ magnitudes in most of the cases. 

The next step was to cross-match all the chips in the 6 different stacked pawprints that create a sequence covering the whole section of a VVV field. For that we used STILTS \citep{tay06}, and allowed a tolerance of $0.34''$, equivalent to 1 pixel-size. We kept all the found objects, and for those detected in more than one image, we adopted their weighted average photometry according to the error reported by DoPHOT. We then cross-matched the two epochs  again from the individual VVV fields in the different filters, again with a tolerance of $0.34''$, i.e., 1 pixel. We kept only sources found in both epochs, and whose differences in magnitudes were less than three times the photometric error provided by DoPHOT. Finally, we cross-matched the photometry from the different filters, and kept sources that show up in at least three filters. This strategy proved to be very successful in eliminating false detections, especially in the proximity of heavily saturated stars. Although DoPHOT masks the inner regions of saturated objects, occasionally it does make some spurious detections in the wings of those sources. Instead of excessively fine-tune some of the input parameters, which could result in losing some real detections, we decided to make use of the fact that we have observations in different filters, and at least a couple of epochs in every filter. Spurious detections in the wings of heavily saturated stars depend on the shape of the saturation and bleeding patterns. These patterns change with time, so spurious detections performed in one epoch are not expected to be repeated in another, and on those few instances where repeated spurious detections are found, they are expected to have very different magnitude values.

\subsection{Catalogs}
\label{sec_catalogs}
\begin{figure*}
   \centering
   \includegraphics[scale=0.5]{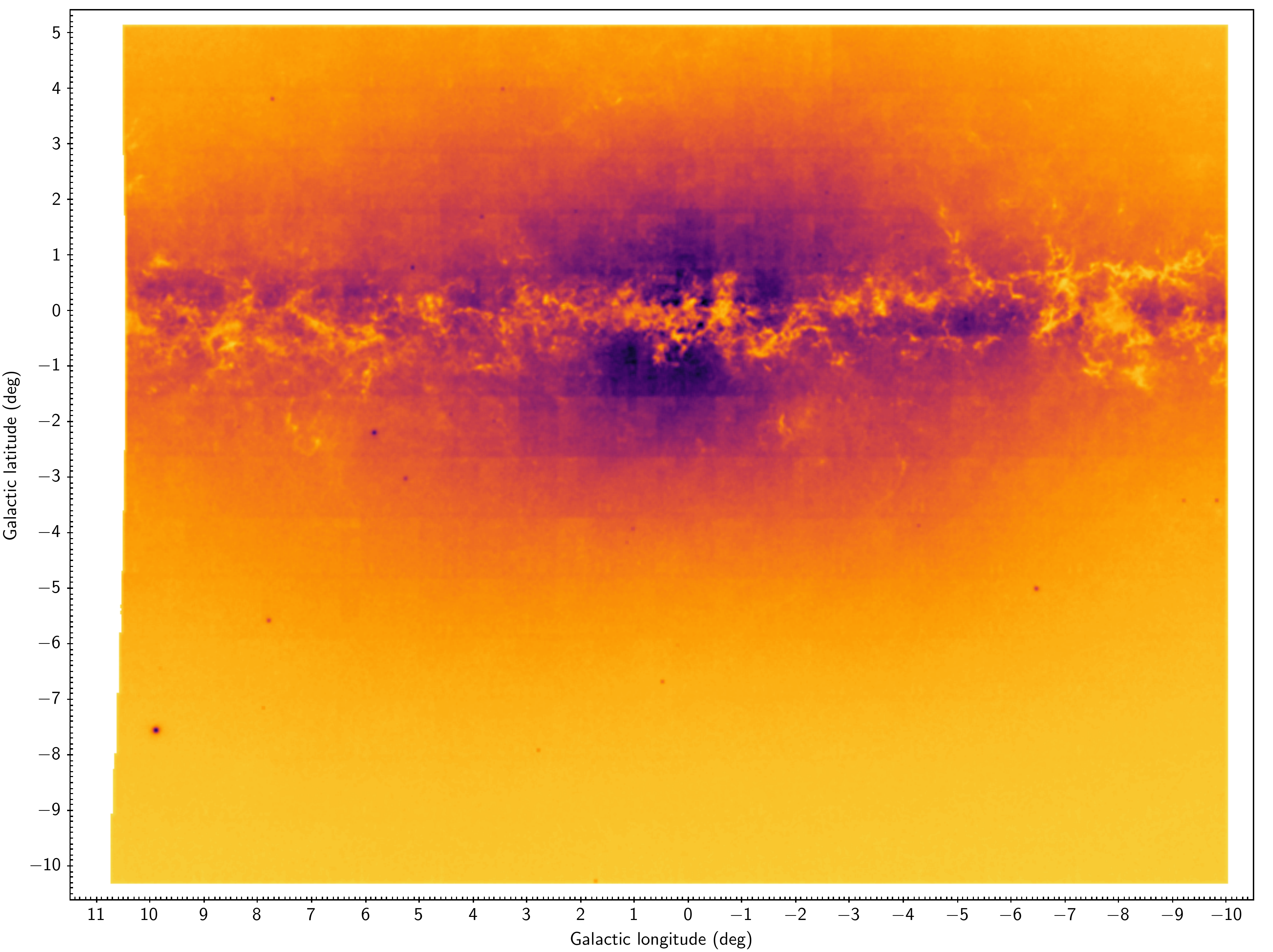}
   \includegraphics[scale=0.33]{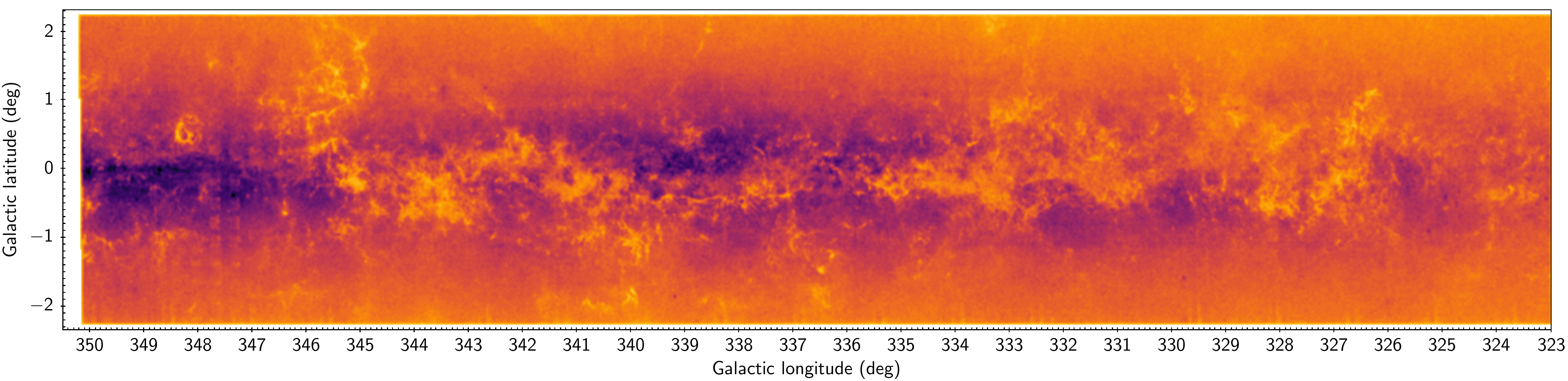}
   \includegraphics[scale=0.33]{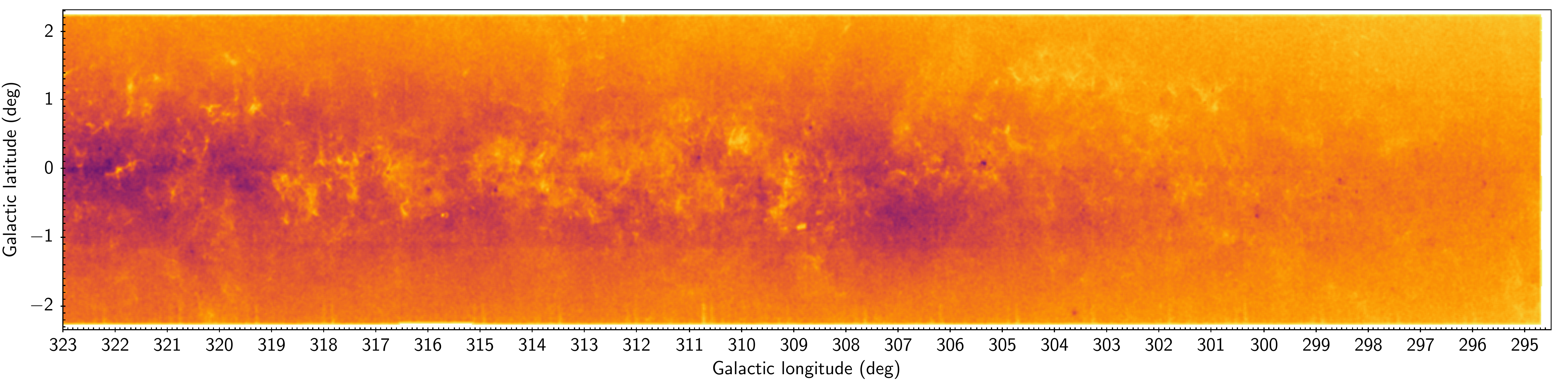}
   \caption{Density maps for all the sources down to $K_s=16.0$ found in the VVV bulge (top panel) and disk (middle and bottom panel) regions. The VVV disk area is split in two panels for clarity. Higher density regions toward the Galactic center and globular clusters are observable in darker colors. Well defined regions of higher extinction in the Galactic plane can be identified in lighter colors. Star clusters appear as darker spots against their lighter-colored surroundings.}
   \label{fig_dens}
\end{figure*}

\begin{figure*}
  \begin{tabular}{ccc}
   \includegraphics[scale=0.38]{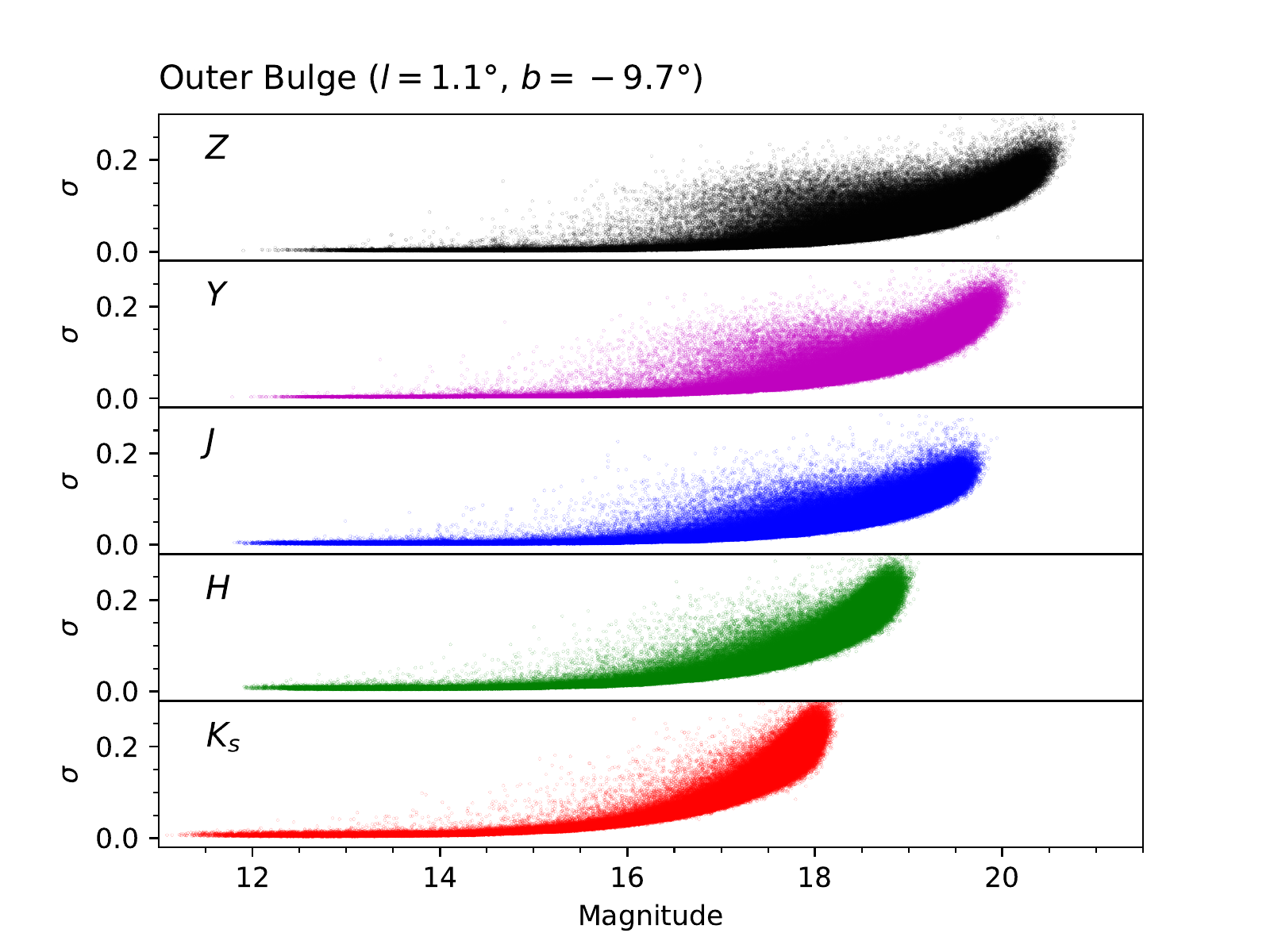}
   \includegraphics[scale=0.38]{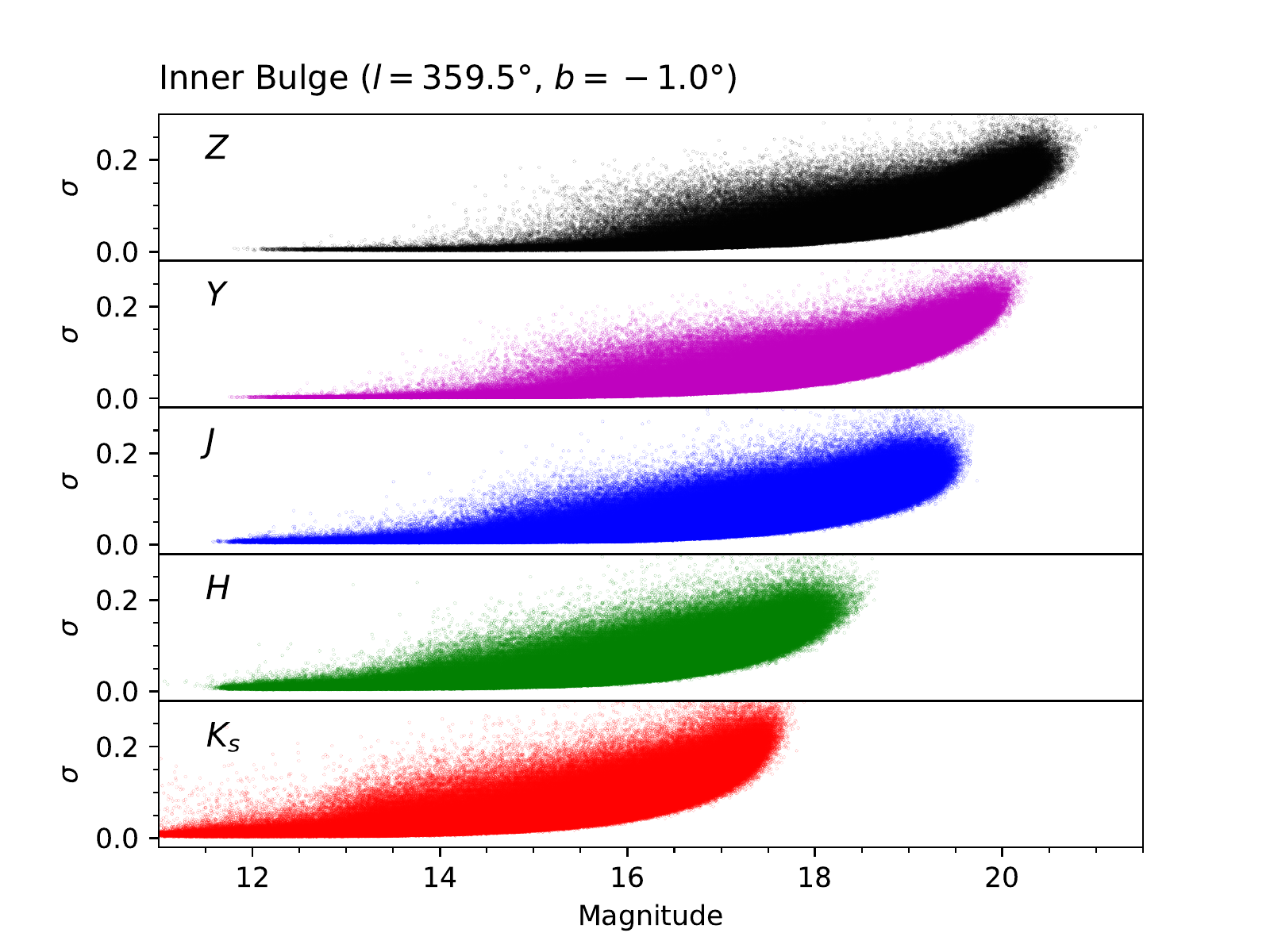}
   \includegraphics[scale=0.38]{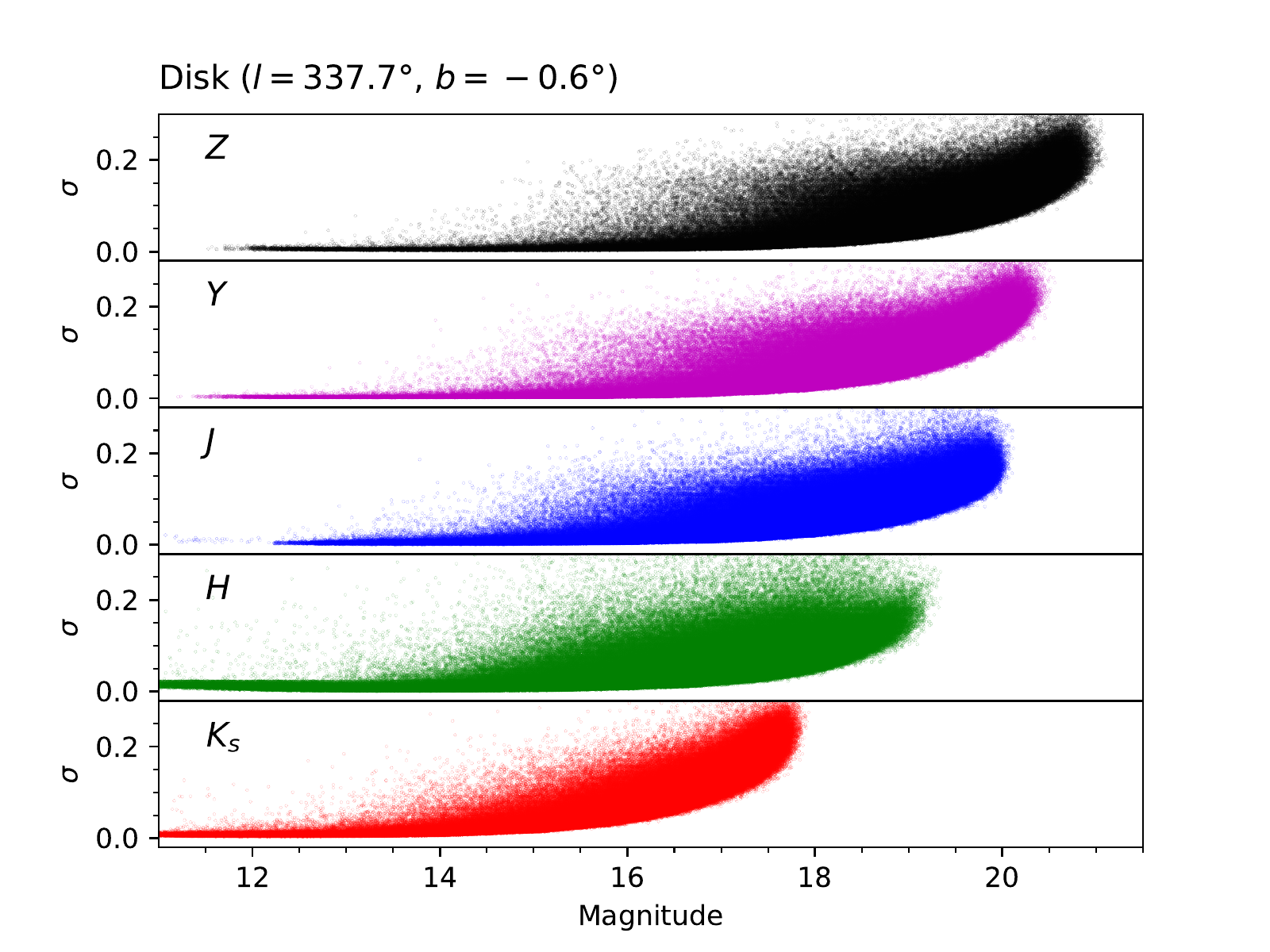}
    \end{tabular}
  \caption{Photometric errors reported in our catalogs for representative fields in the VVV outermost bulge (left panel, at VVV field b208, centered at $l=1\fdg1$, $b=-9\fdg7$), VVV innermost bulge (middle panel, at VVV field b319, centered at $l=359\fdg5$, $b=-1\fdg0$) and VVV disk (right panel, at VVV field d068, centered at $l=337\fdg7$, $b=-0\fdg6$), for the different near-infrared filters used. The depicted magnitudes and photometric errors are averages of the individual measurements in the different stacked images in the different epochs used, weighted according the photometric errors reported by DoPHOT.}
   \label{fig_photerror}
\end{figure*}

\begin{figure*}
  \begin{tabular}{ccc}
   \includegraphics[scale=0.27]{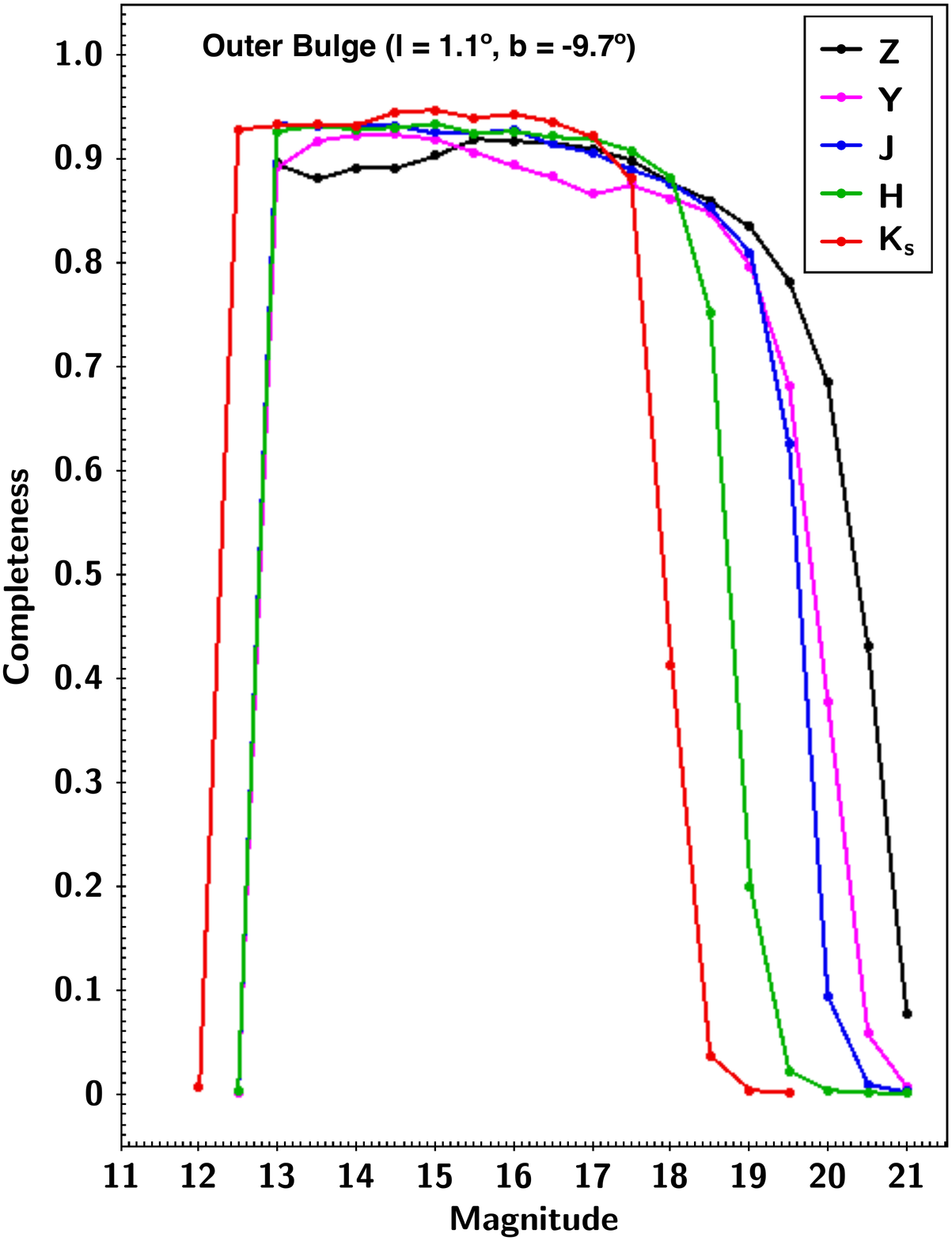}
   \includegraphics[scale=0.27]{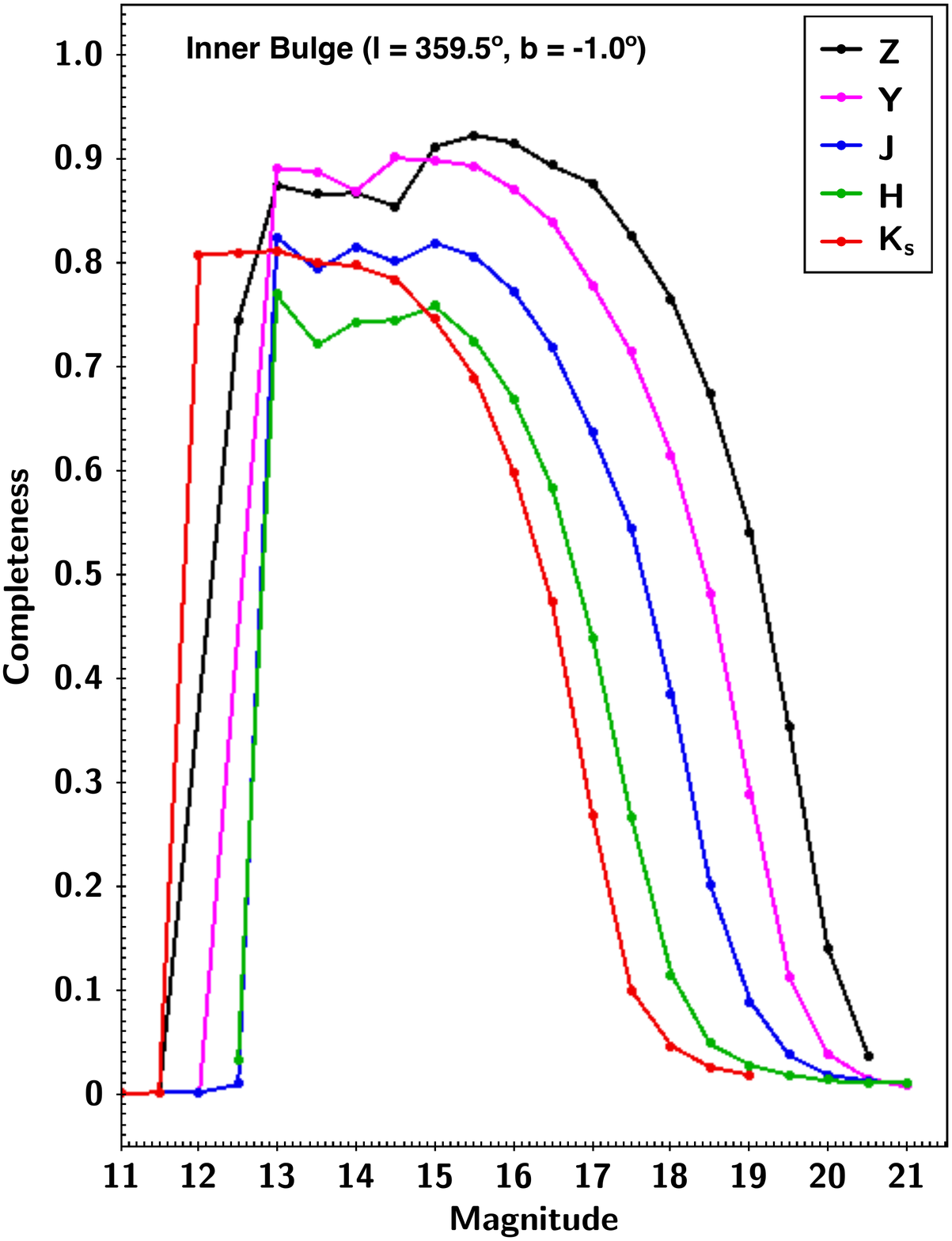}
   \includegraphics[scale=0.27]{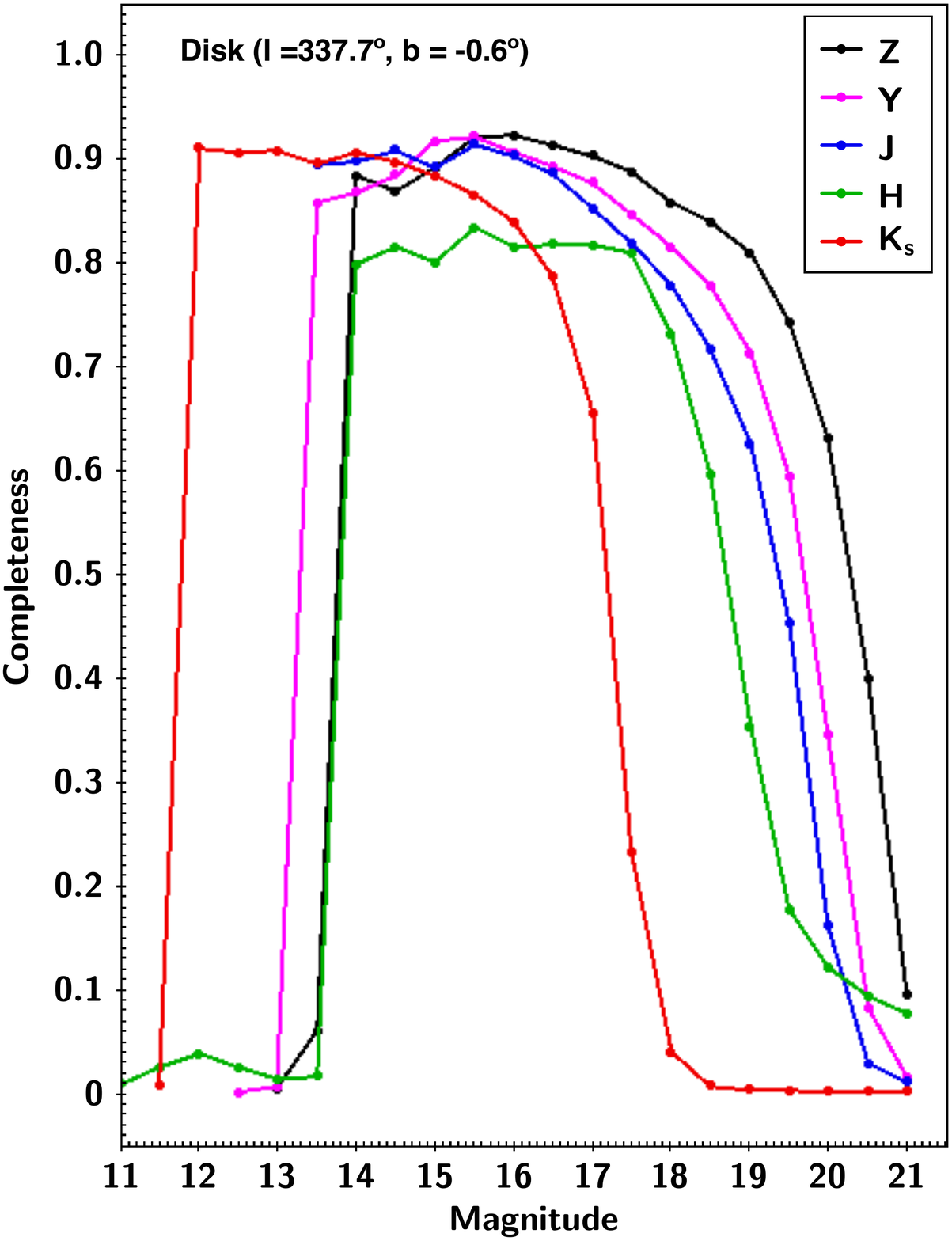}
    \end{tabular}
   \caption{Completeness profiles for representative fields in the VVV outermost bulge (left panel), VVV innermost bulge (middle panel) and VVV disk (right panel). The analyzed VVV fields are the same as in Figure~\ref{fig_photerror}, but in this case only one chip was used per field. As expected, the innermost regions, more heavily affected by extinction, show lower completeness levels. The disk area, although also affected by high extinction, is not so affected by crowding, leading to higher completeness levels than in the innermost bulge.}
   \label{fig_comp}
\end{figure*}

\begin{figure*}
  \begin{tabular}{ccc}
   \includegraphics[scale=0.27]{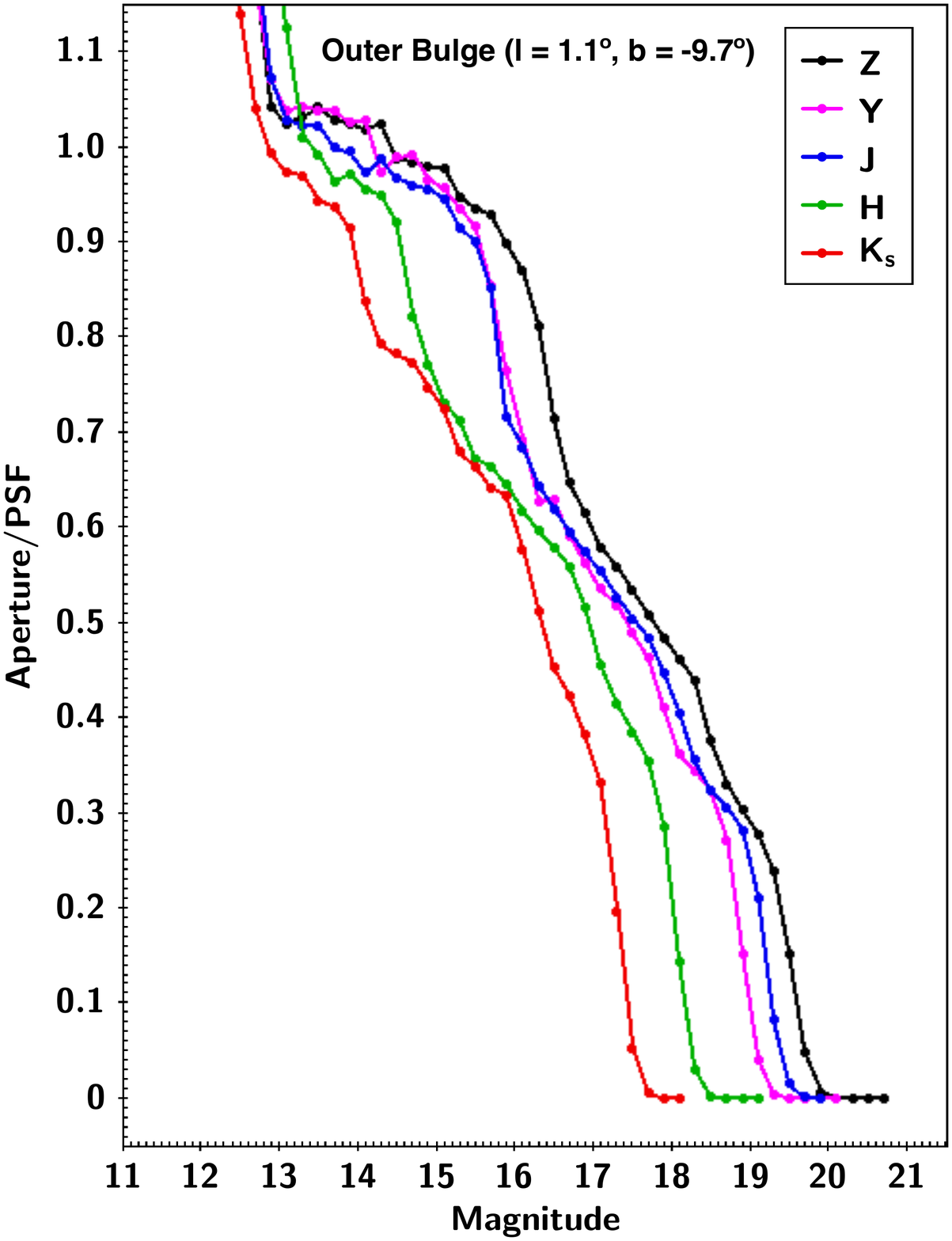}
   \includegraphics[scale=0.27]{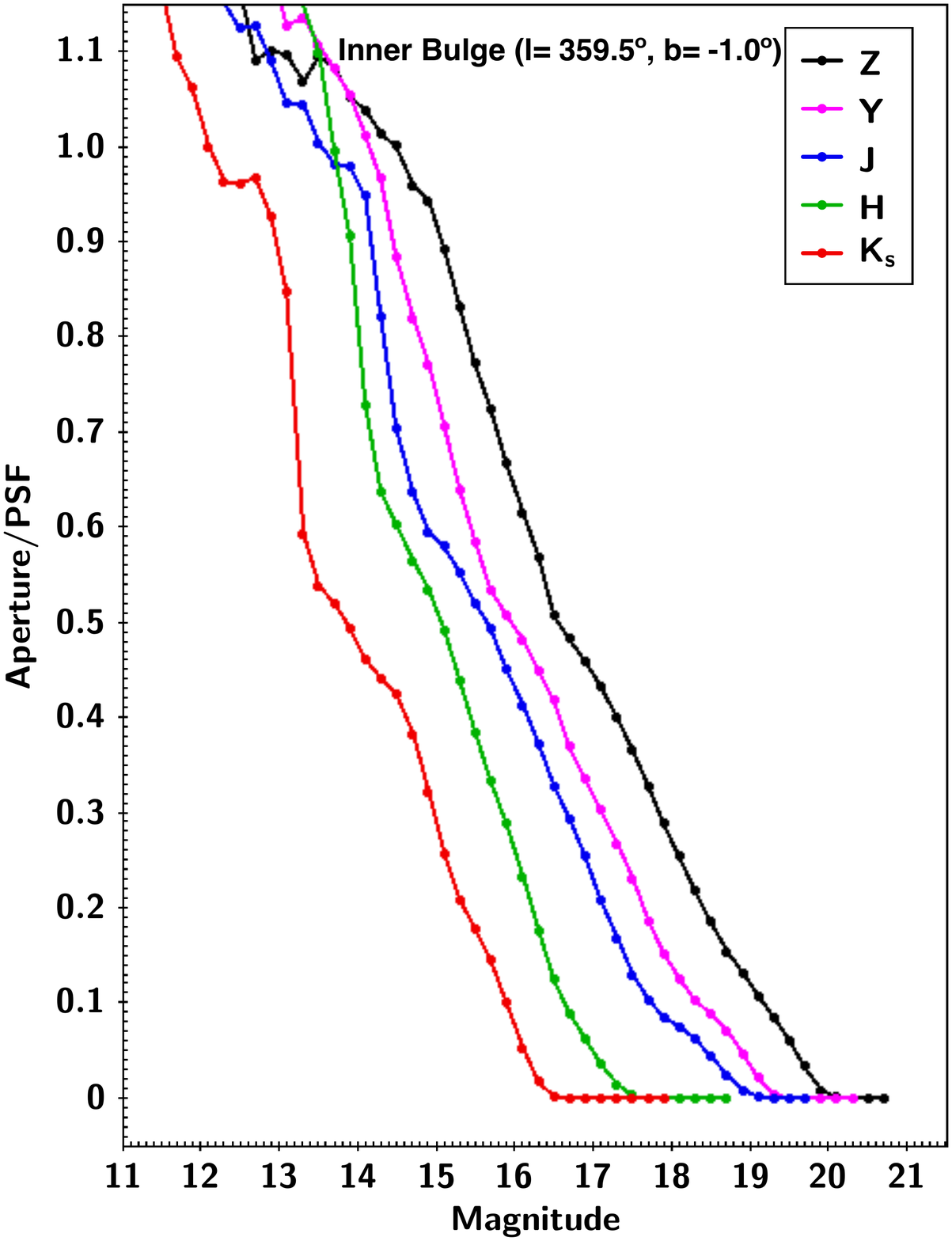}
   \includegraphics[scale=0.27]{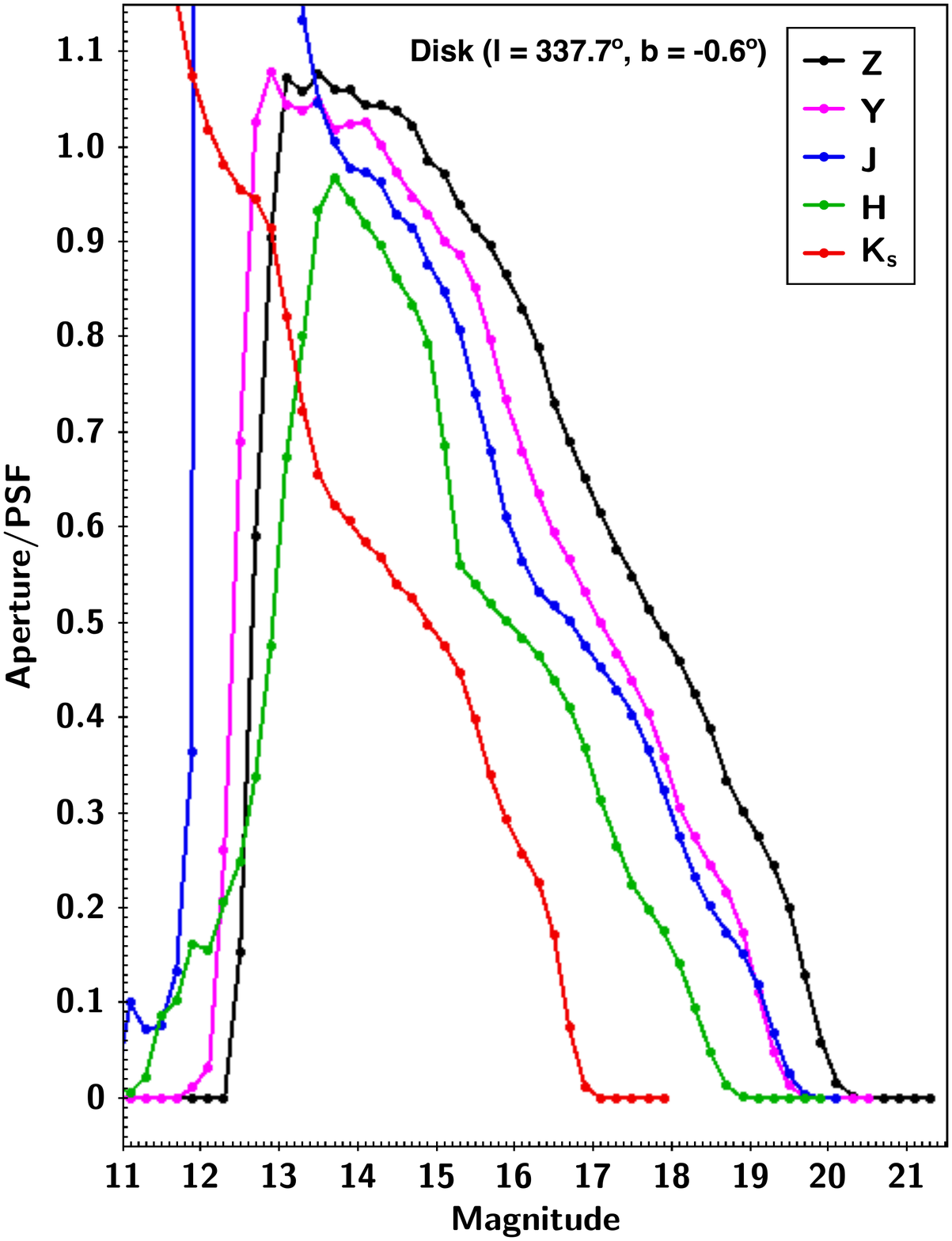}
    \end{tabular}
   \caption{Ratios of the number of detected sources as a function of magnitude between our PSF photometry and CASU aperture photometry catalogs for representative fields in the VVV outermost bulge (left panel), VVV innermost bulge (middle panel) and VVV disk (right panel). The analyzed VVV fields are the same as in Figure \ref{fig_photerror}. Our PSF photometry reaches dimmer detection limits, by as much as $\sim1-1.5$ for the most crowded region in the inner bulge. The completeness is also higher over most of the magnitude range for all the different filters, and in all the different regions surveyed, while maintaining small photometric errors (see Figure \ref{fig_photerror}). The higher detection ratios with aperture photometry at the bright end are due to the use of a more conservative saturation limit with the PSF photometry.}
   \label{fig_comp2}
\end{figure*}

In the end, we were able to detect and provide photometry for 846 million sources in our final catalogs from the inner Galactic regions observed by the VVV survey (570 millions in the bulge area, and 276 millions in the disk area). We plot them in Figure \ref{fig_dens} to check their distribution. As previously seen in \citet{sai12a} and \citet{sot13}, densities increase toward lower latitudes, although higher extinction reverses this tendency at latitudes around $|b|\le1\degr$. Galactic star clusters located in the VVV area reveal themselves in Figure \ref{fig_dens} as darker spots due to their increased stellar densities. In Figure \ref{fig_photerror}, we can observe the high quality of the photometry reported in our catalogs, and how the position in the inner Galaxy affects it, as expected by the significant changes in crowding along the area surveyed by VVV. To provide a measurement of the completeness of the extracted PSF photometry, we performed artificial star tests in three representative fields in the VVV area: one in the outer bulge, one in the inner bulge, and one in the disk (see Figure \ref{fig_comp}). For a given artificial star test, we injected a different set of 5,000 sources, well spread all over the image, all with the same magnitude. We repeated this test numerous times, changing every time the magnitudes of the injected stars in a range between 10 and 21, in intervals of 0.5 magnitudes, for the five different filters, and for the two used epochs per filter. To speed up the process, only one of the 16 detectors of the camera was used. Since we only injected 5,000 sources at a time, spread all over the image, we did not significantly alter the crowding of the image, while having enough sources to obtain good statistics for our tests. For the injected source to be reported as recovered, it has to be measured in both epochs for a given filter, and with a recovered magnitude that has to be within 3 times the dispersion measured for the recovered sample at that given magnitude and filter, when averaged over both epochs. We find that while in the less crowded regions the completeness is very high, over $90\%$ on a five-magnitude interval, the most crowded regions get lower completeness rates, between $80\%$ and $90\%$ on intervals of three to four magnitudes, quickly decreasing thereafter. However, in Figure \ref{fig_comp2} we can see that the completeness level in our catalogs is always significantly higher than in the catalogs obtained using aperture photometry available at the ESO archive. The artificial star tests also allowed us to calculate the dispersion between injected and recovered magnitudes of stars and compare them with the reported errors in our photometry and shown in Figure \ref{fig_photerror}. We see the agreement between both to be dependent on the level of crowding. We found that the reported photometric errors in our catalogs are similar to the ones in the artificial star tests for the outer regions of the bulge, while they seem to be underestimated by a factor of two with respect to the ones from the artificial star tests for the inner regions of the bulge. In the disk region they seem also to be underestimated, but by a smaller factor.

The catalogs with the PSF photometry for all the area surveyed by the VVV in the five near-infrared ZYJHKs are publicly available for the whole community through the VISTA Science Archive (VSA)\footnote{http://surveys.roe.ac.uk/vsa}, ingested into two tables linked to the other VVV data. These tables are \verb+vvvPsfDophotZYJHKsMergeLog+ and \verb+vvvPsfDophotZYJHKsSource+, designed in the similar style to \verb+vvvMergeLog+ and \verb+vvvSource+ and other VDFS band-merged tables \citep{ham08,cro12}. The mergelog links each of the 348 VVV fields (196 for the bulge and 152 for the disk), to the two epochs of ZYJHKs tile multiframes and the  \verb+vvvPsfDophotZYJHKsSource+ includes the photometry and includes a {\it priOrSec} attribute that allows selection of a seamless catalog across all fields. Furthermore neighbor tables  \citep{ham08,cro12} link \verb+vvvPsfDophotZYJHKsSource+ to the band-merged aperture photometry table \verb+vvvSource+, (\verb+vvvSourceXPsfDophotZYJHKsSource+) and to external surveys, e.g. VPHAS+, ALLWISE, GLIMPSE, GaiaDR1,etc. (e.g. \verb+vvvPsfDophotZYJHKsSourceXglimpse2_hrc+). The contents of the tables and list of neighbor tables can be found in SchemaBrowser\footnote{http://surveys.roe.ac.uk/vsa/www/vsa\_browser.html}, following the index to VSA VVV, VVVDR4, and Tables. The default output for a search using the \verb+vvvPsfDophotZYJHKsSource+ table includes 13 different columns: name of the VVV field the object was detected in, right ascension, declination, and magnitude and photometric error in the five available near-infrared filters ($ZYJHK_s$). The reported right ascension and declination are averages over the different epochs in the different filters, while the reported magnitude and errors are weighted averages, as a function of the photometric error reported by DoPHOT, of the magnitudes and errors reported in the different epochs used in a given filter. The total number of detected sources in the catalogs changes significantly within the different regions of the sky surveyed by VVV. In the bulge area, it varies from the outer, less-populated regions of the bulge, with nearly a million detections per field, up to the inner regions where detections go up to more than five million sources in some fields. In the disk surveyed area, detections change from one million sources per field for the regions far away from the bulge, to three million sources per field for regions closer to it. 

\subsection{Caveats in the procedure}
\label{sec_caveats}
Although the strategy described in this section to produce our final cross-matched photometry allows for the creation of very clean CMDs shown in Section \ref{sec_cmd}, there are some caveats that potential users should be aware of when planning to work with our catalogs.

Our self-imposed rules to keep only sources appearing in two epochs in at least three filters proved to be very successful in getting rid of most spurious detections. An inevitable consequence of this method, however, is the omission in our final catalogs of real variable sources, but only if the observations were done when the difference in phase between the epochs of the observations of a given source results in a magnitude variation for that source that is above our implemented tolerance, that is, three times the photometric error.

Users should be aware of the absence in our final catalog of high-velocity objects. Sources do not appear in the catalog if they move more than our cross-match tolerance ($0.34\arcsec$, i.e., 1 pixel) within the analyzed images, either in the ones for the different epochs in a given filter or in the ones from different filters.
 
\begin{figure*}
  \begin{tabular}{ccc}
    \includegraphics[scale=0.33]{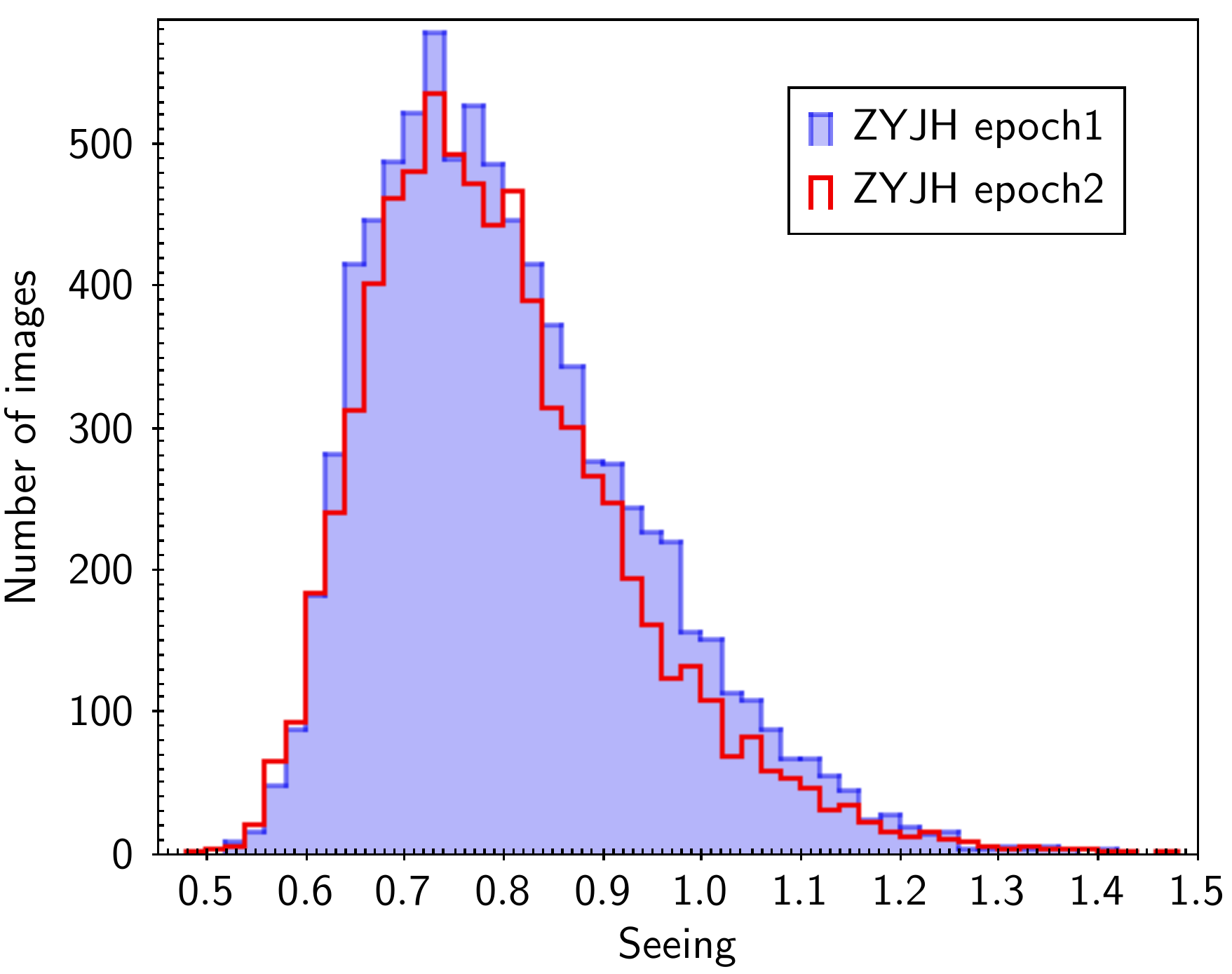} &
    \includegraphics[scale=0.33]{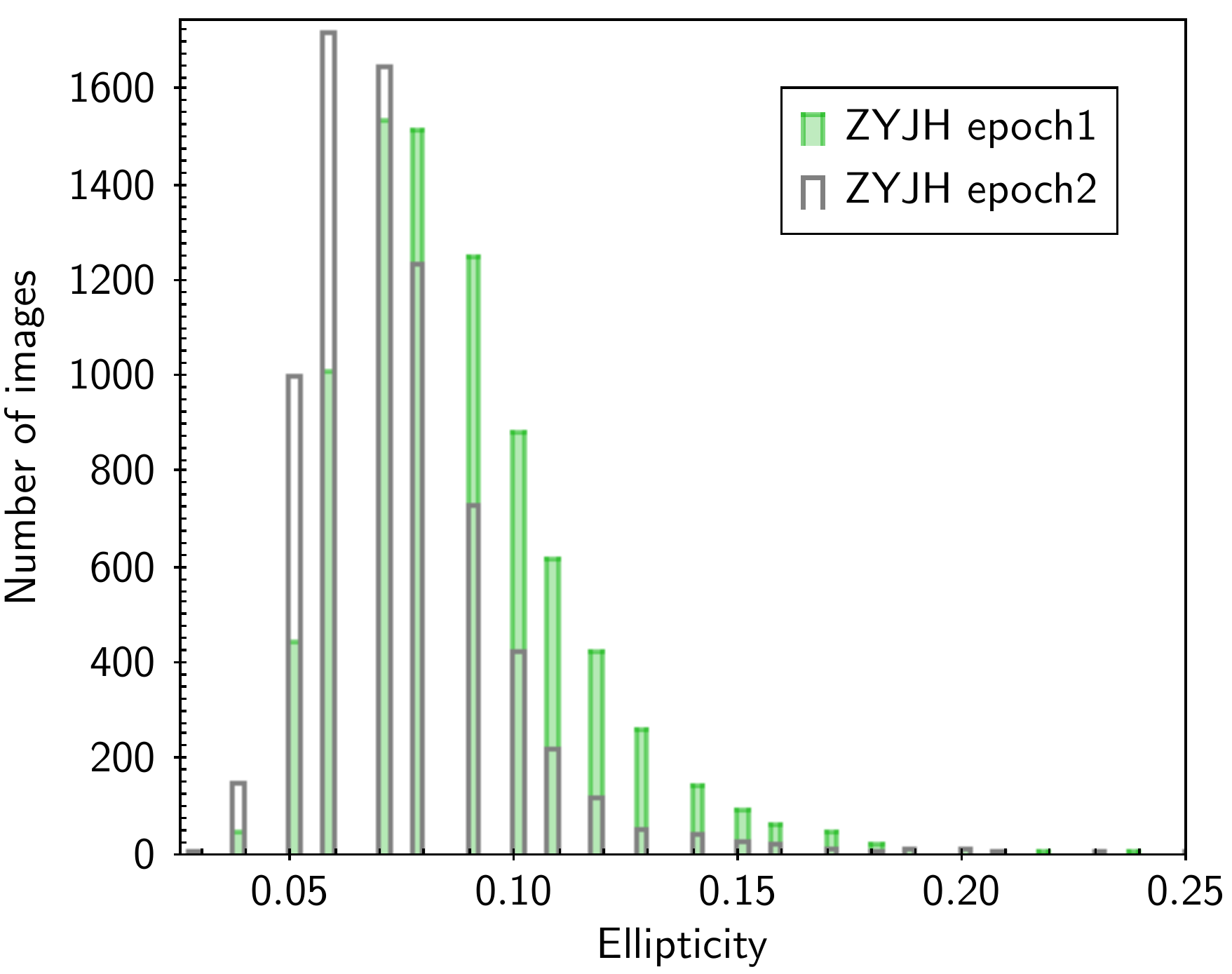} &
    \includegraphics[scale=0.33]{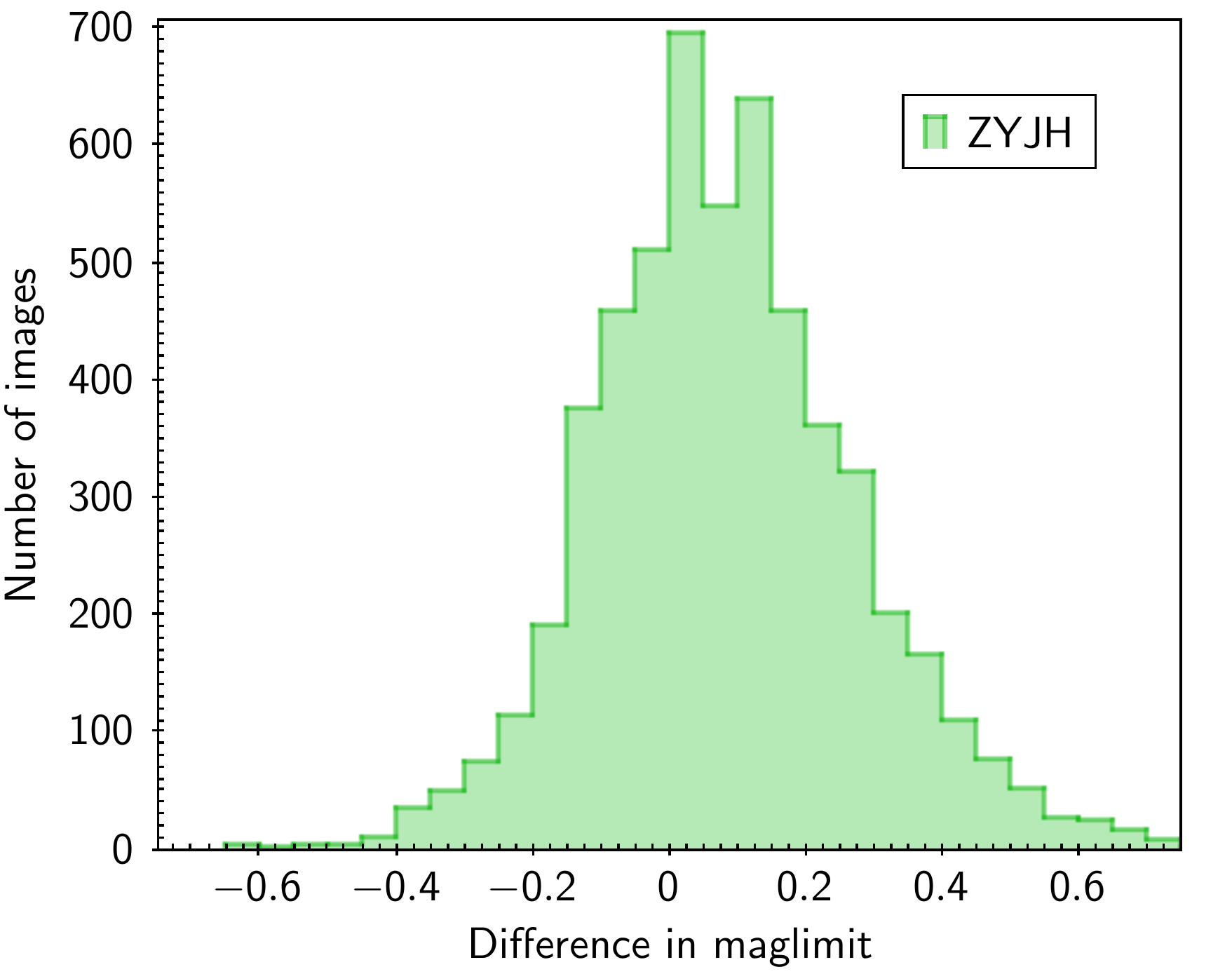} \\
    \includegraphics[scale=0.33]{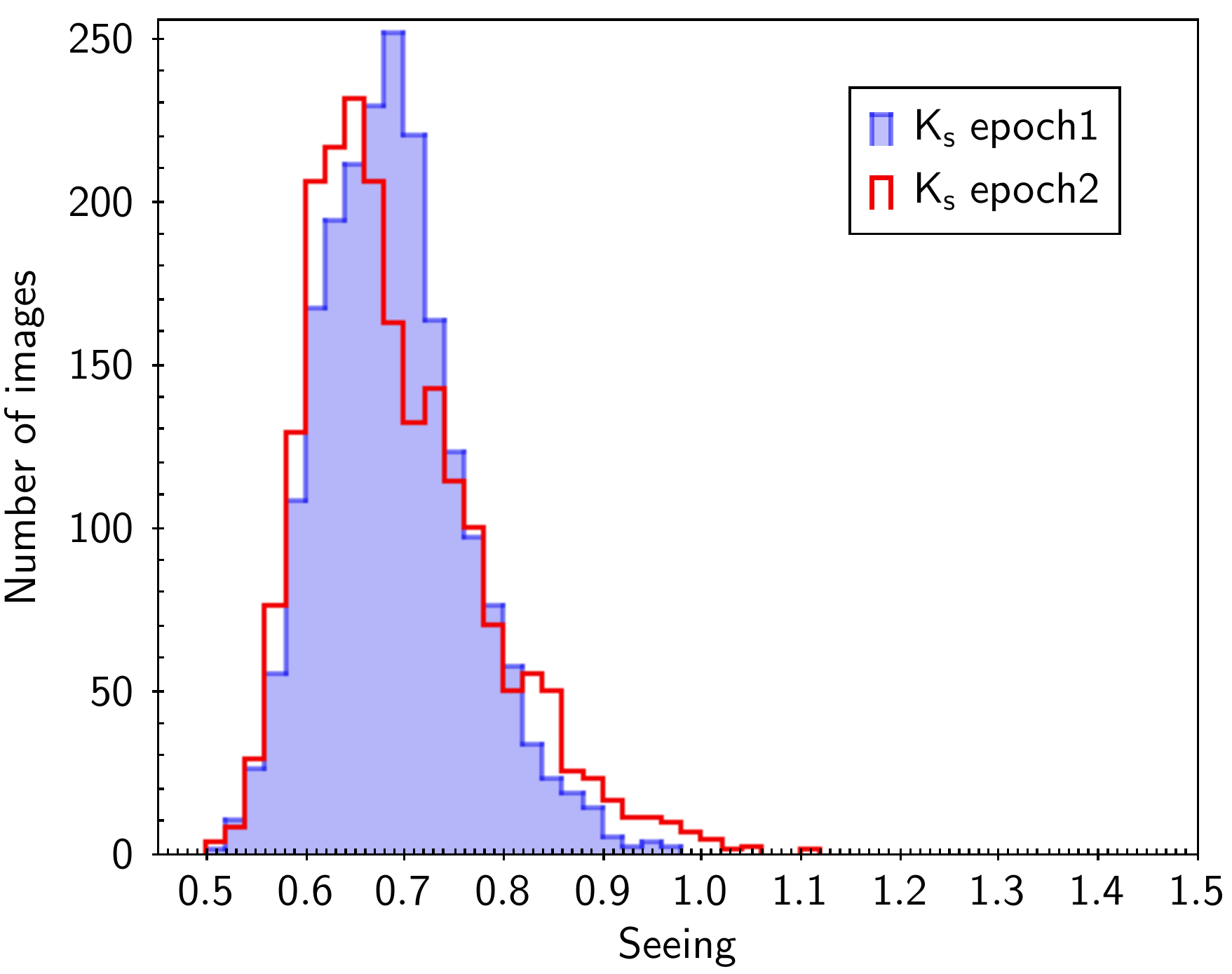} &
    \includegraphics[scale=0.33]{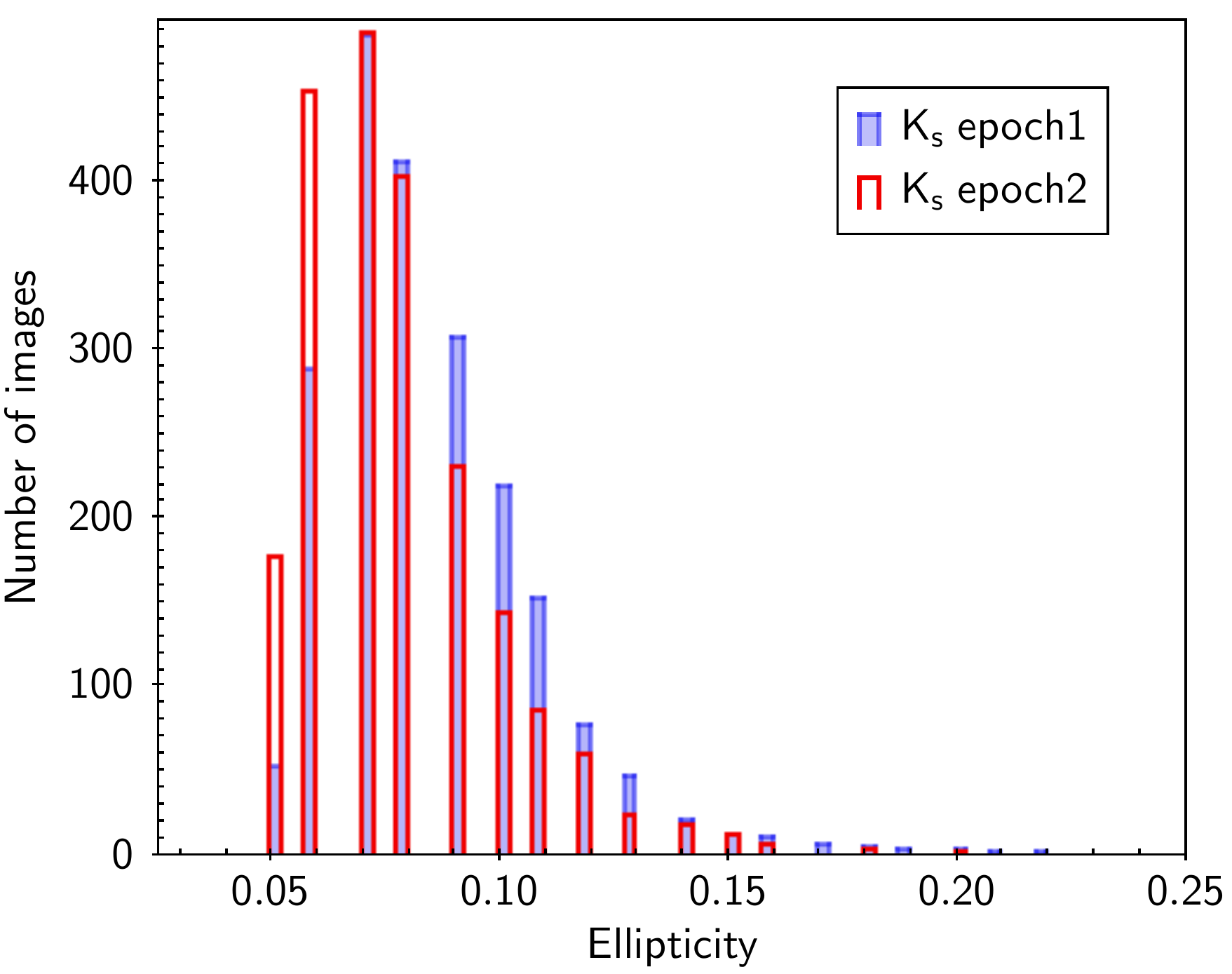} &
    \includegraphics[scale=0.33]{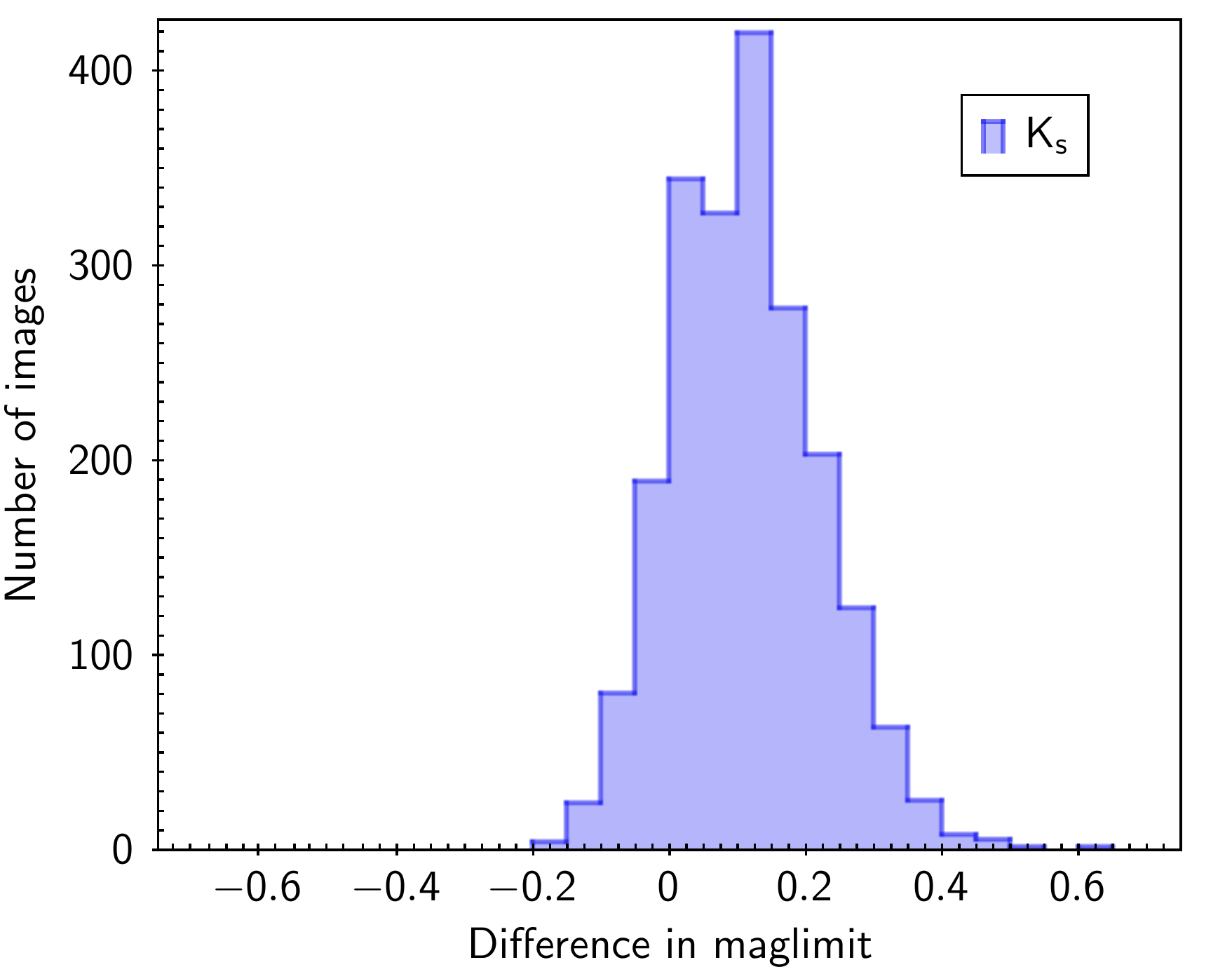} \\
    \end{tabular}
  \caption{Distribution of the parameters (seeing, ellipticity and magnitude limit per stack image) assessing the high quality of the VVV images used for obtaining our new photometric dataset. We show the $ZYJH$ filters (upper panels) and the $K_s$ filter (lower panels). In the left and central panels, the green and blue filled histograms correspond to epoch 1, while the gray and red empty histograms correspond to epoch 2. In the right panels, the difference between the magnitude limits is in the sense of first epoch minus second epoch. In the upper right panel, we have omitted from the comparison the images from the disk in $Z$ and $Y$ due to the different exposure times between epochs (see Table~\ref{tab_exptimes}). We note that the higher quality of the $K_s$ images on average is due to having more images to select from.}
   \label{fig_quality}
\end{figure*}

Users should also realize that the different atmospheric and telescope conditions during the two epochs make the seeing and the detection limit change between both periods used for the analysis in a given filter. Although observing conditions were always very good, with clear skies and usually sub-arcsecond seeing conditions (see Figure \ref{fig_quality}), observing restrictions were more stringent in the 2010-2011 period (epoch 1) than in the 2015 period (epoch 2) for $J$ and $H$ images. This fact, together with a light sensitivity degradation over time, meant a better detection limit on average in epoch 1 (see right panels in Figure \ref{fig_quality}), and the absence in our final catalog of the dimmest objects found in the epoch with a deepest detection limit.

Additionally, when using catalogs of adjacent fields users should take into account the presence of repeated sources. We decided to release the photometry according to the VVV fields the different sources were detected on (see Section \ref{sec_catalogs}). Users should be aware however that there exists a small overlap among the different VVV fields \citep{sai12a}, which implies that a small percentage ($\sim6\%$) of the reported sources in a given field are also found in another one. Duplicate sources in a search in the catalog are easily identified through the {\it priOrSec} attribute.

Finally, we would like to mention that, in a very few cases, observations in one epoch had a problem, for example, observation done out of focus, or while the telescope was still moving. When, for a given epoch and filter, the average seeing on the chips in one stacked pawprint was larger than $1.5''$ and the average ellipticity of the sources was over $0.25$, that epoch was not considered. This results in a small percentage of fields ($\sim2\%$ in $Z$, $Y$, $J$ and $H$) with only one epoch available in a given filter. We also faced the problem of very high sky levels on average on epoch 2 in the $H$ filter observations in the disk region, which decreased considerably the dynamic range for the detections with respect to epoch 1. This was probably produced by the combined effect of higher sky values at longer wavelengths, longer exposure times for H images in the disk regions (see Table \ref{tab_exptimes}), and exposures taken for this particular filter and epoch close to twilight in a significant number of VVV disk fields. For this particular filter and region we decided to consider only observations in one epoch.    

\section{The VVV giga-CMD}
\label{sec_cmd}
Previous articles in this series already provided a first look and analysis of the CMDs of VVV in the areas surveyed in the bulge \citep{sai12a} and in the disk \citep{sot13}. However, our new PSF photometry allows us to build CMDs that go deeper and are significantly more complete than previous ones, leading to a better definition and understanding of the stellar populations present in the inner regions of the Galaxy. Considering the wide-field coverage of the internal regions of our Galaxy by VVV, we reckon that plotting a single CMD with all the stars detected does not provide a clear view of the stellar populations residing in that region. We prefer to provide CMDs of smaller subregions and analyze them qualitatively, comparing the striking differences that we can observe between the CMDs of some of the studied subareas. We do this in the next subsections, but to obtain a complete view of all the surveyed regions, we provide a series of videos that accompany this article. In these videos, we show the different CMDs for all the different fields and filters available in the VVV survey. 

\subsection{CMDs of the Galactic bulge fields in the VVV}
\label{sec_cmdBulge}
\begin{figure*}
   \centering
   \includegraphics[scale=0.85]{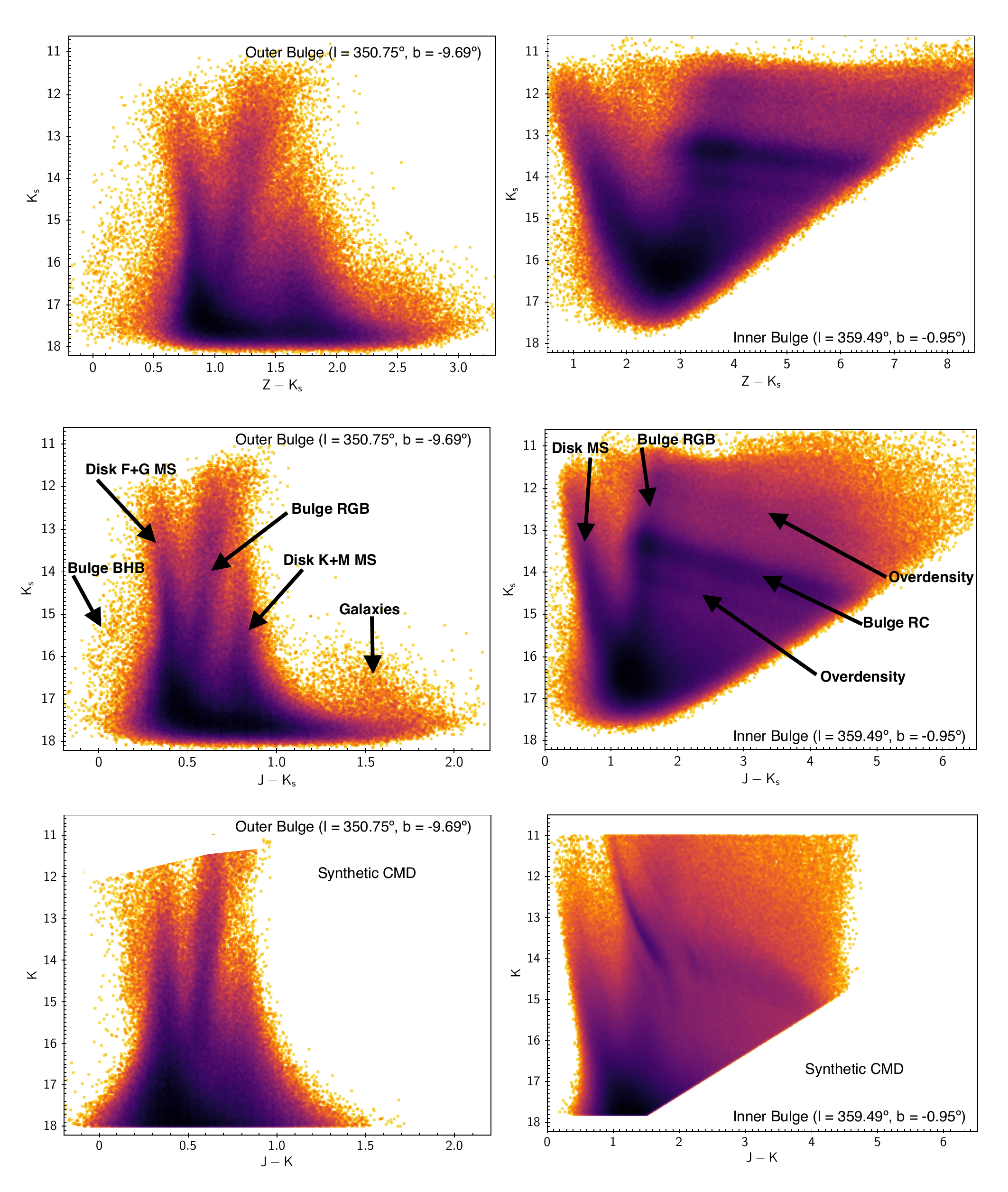}
   \caption{CMDs of Galactic bulge regions for VVV representative fields at the uppermost (left panels, at VVV field b201, centered at $l=350\fdg75$, $b=-9\fdg69$ in the outer bulge) and lowermost (right panels, at VVV field b319, centered at $l=359\fdg49$, $b=-0\fdg95$ in the inner bulge) latitudes available to the VVV observations. The upper panels represent the $Z-K_s$ vs $K_s$ CMDs, while the middle panels represent the $J-K_s$ vs $K_s$ CMDs build from our PSF photometry. Darker colors represent higher densities of sources. In the lower panels, synthetic CMDs built using Besan\c{c}on Galactic model \citep{rob03}. In the outer bulge CMD we are able to separate the different sequences of stellar populations mentioned in the text. In the inner bulge CMDs those sequences are significantly blurred out due to differential extinction, but some overdensities from the bulge RGB manifest clearly as described in the text. A series of online movies show the CMDs for the different VVV bulge fields.}
   \label{fig_cmdBulge}
\end{figure*}

The CMDs of the Galactic bulge area covered by the VVV survey change dramatically with latitude by the effect of the huge variation in interstellar extinction. Even though VVV is a near-infrared survey, the effects of interstellar extinction in the CMD are clearly shown in the accompanying videos and in Figure \ref{fig_cmdBulge}.

In the left panels of Figure \ref{fig_cmdBulge}, we show the CMDs of a field in the outermost VVV bulge area, only affected by very mild extinction in the near-infrared. The CMD shows the three conspicuous stellar branches described in \citet{sai12a}: on the blue edge, we can distinguish the main sequence (MS) of disk F and G stars in front of the bulge, slightly redder are the red giant branch (RGB) bulge stars, and even redder the MS disk K and M stars. In addition we are also able to identify other two groups of sources, one in the extreme blue region that we identify as blue horizontal branch (BHB) of the bulge, and the other in the extreme red dimmest region of the CMDs, that we identify as background galaxies.

The disk MS of the F and G stars extends from the brighter magnitudes available from our photometry ($K_s\sim12$) in the blue ($J-K_s\sim0.2$ and $Z-K_s\sim0.6$), down to our faintest magnitudes ($K_s\sim18$) and moving toward redder colors ($J-K_s\sim0.5$ and $Z-K_S\sim1.0$). Down to magnitude $K_s\sim16$ this sequence is well separated, whereas at fainter magnitudes it becomes more and more contaminated with bulge stars. The stars from the bulge extend also from the brightest sources available from our photometry ($K_s\sim11.5$) a little more into redder colors than the disk MS of F and G stars ($J-K_s\sim0.7$ and $Z-K_s\sim1.4$) down to the faintest magnitudes ($K_s\sim18$) and moving toward bluer colors ($J-K_s\sim0.5$ and $Z-K_s\sim1.0$). The RGB of the bulge is observable as an independent branch down to its base at magnitudes $K_s\sim16$, but the dimmer sub-giant branch (SGB) and upper MS of the bulge get mixed with the foreground MS of the disk. Starting at the brightest magnitude magnitudes available in our CMDs confused with the bulge RGB, but becoming a clear independent sequence at $K_s\sim14$, and descending to our faintest available magnitudes almost vertically at colors $J-K_s\sim0.85$ ($Z-K_s\sim1.7$), we can find the sequence of the disk MS K and M nearby dwarf stars in the foreground of the bulge. As detailed in \citet{luc08}, the reason for the two-disk MS is related to the combination of the effects produced by differences in distances among the MS disk stars, the colors of different types of the stars present and their relative number. Late K and M stars have a short color range at near-infrared wavelengths ($J-K_s\sim0.9$) that clumps them in a well-defined sequence. Those that we are able to observe are the closest ones to the Sun, and as we move away to further galactic distances, this type of star becomes too faint to be detected. On the other hand, we can detect F and G stars that are further away in the disk, and that were saturating before. These are the stars that populate the bluer disk MS. The apparent separation between sequences is the result of the late G and early K type stars to have a higher color spread in the near-infrared CMDs.

The blue stars located at colors $J-K_S\sim0.2$ ($Z-K_s\sim0.4$) and magnitudes $K_s<14$ down to $K_s\sim17$ or even more, and $J-K_S\sim0$ ($Z-K_s\sim-0.1$), correspond to the BHB stars belonging to the bulge. BHB stars are low-mass, core-helium burning stars, that are located to the right of the RR~Lyrae instability strip (e.g.,\citet{cat15}, and references within). They are old objects, usually used to trace old, metal-poor structures such as the Galactic halo \citep{cat09a,vic12}. In the Galactic bulge, although metal-rich BHB stars exist, the vast majority of them seem to be metal-poor \citep{pet01,ter04}. Identifying old metal-poor stars in the fields of the Galactic bulge is difficult, as they make up a small fraction compared to the metal-rich component. But they are an intrinsically interesting population, since different simulations show that they can be the oldest stars in the Galaxy. There are some ongoing efforts to look for these metal-poor stars, but observed targets are in the relatively few inner metal-poor globular clusters \citep{bar14}, or on small samples selected from large spectroscopic surveys \citep{siq16}. The identification of a significant amount of BHB stars in the VVV bulge fields presents a way to improve the tracing of the old metal-poor population of stars in the inner Galaxy, and in \citet{mon18} we present an atlas of several thousands of VVV BHB stars. They are certainly more numerous than the other tracer used so far to map the metal-poor component of the inner regions of the Galaxy, the RR~Lyrae \citep{dek13,min16}. Although their faint magnitudes are a handicap for spectroscopic follow-up,  they represent an interesting target for upcoming spectroscopic surveys with the next-generation instruments in 8m-class telescopes, such as the Multi-Object Optical and Near-IR Spectrograph (MOONS).

Another interesting feature of these CMDs is the branch of very dim and red sources that appears at magnitudes fainter than  $K_s>16$ and colors $J-K_s>1.3$ ($Z-K_s>2.3$). These sources can be associated with background galaxies, as some studies suggest \citep{luc08,col14,bar18}, although at the faintest magnitudes this sequence is expected to be heavily contaminated with the tail of Galactic stars with higher photometric errors. We note that this sequence is more noticeable in the $K_s$ vs $J-K_s$ CMD, while the previously mentioned BHB sequence manifests more clearly in the $K_s$ vs $Z-K_s$ CMD.

The synthetic CMD built using the Besan\c{c}on Galactic model \citep{rob03} shown in the lower left panel in Figure \ref{fig_cmdBulge} reliably represents the  most populated sequences, that is, the MS disk stars, RGB bulge stars and K and M dwarf stars. But significantly, it fails to reproduce the BHB sequence, and, as expected since it is a Galactic model, it does not show the sequence of background galaxies.

The features of the CMD get significantly blurred out by interstellar extinction when we go to latitudes closer to the Galactic plane. In the right panels of Figure \ref{fig_cmdBulge}, we notice that all branches in the CMD move toward redder colors when compared with the CMD in the left panel of the outer bulge region. We can still appreciate on the blue side of the CMD ($J-K_s<1$) the stars in the MS of the disk, and to the red ($J-K_s>1$) the RGB of the bulge. It is immediately obvious that the RGB is more populated now, as expected since we are moving to lower latitudes and more stars from the bulge are in our line of sight. It is also discernible that the RGB extends several magnitudes in color, caused by the effect of the severe differential extinction present \citep{alo17}.

There are some overdensities that are also quite clear now in the RGB of the Galactic bulge. The most distinct is the one associated with the red clump (RC) of the metal-rich stars in the Galactic bulge. It is located in the region shown in the CMD in the right panel at $K_s\sim13.5$ and $J-K_s\sim1.6$ ($Z-K_s\sim3.5$) and extends all the way down to $K_s\sim14.5$ and $J-K_s\sim4.5$ (in the upper panel of Figure \ref{fig_cmdBulge}, the limit in magnitude is $K_s\sim14.0$ at $Z-K_s\sim6.5$ due to the $Z$ magnitude limit). The RC of the bulge has been used to learn about the morphology and physical parameters of the inner Galaxy due to its characteristics as standard candle. Using 2MASS and previous VVV  near-infrared photometry, \citet{sai11,sai12a} have shown the presence of a double RC at latitudes $b<-8\degr$, which they associate to the Galactic bulge having an X-shape. Using our new photometry, \citet{gon15} have reinforced the link between the double RC and the X-shape of the Galactic bulge. Using information of the RC stars of the Galactic bulge from our new VVV PSF photometry, \citet{val16} have first provided the RC density map of the bulge across the VVV area, then used it to properly scale the initial mass function measured in a small {\em Hubble Space Telescope} bulge field, and provide a fully empirical estimate of the bulge stellar mass. The RC of the bulge from our new PSF photometry has also been used to learn about the extinction toward the inner Galaxy. \citet{alo17} have provided a new measure of the extinction law toward the innermost bulge, defining the reddening vector as described by the RC stars. \citet{nat16} have also used our new VVV photometry, along with OGLE optical photometry, to follow the RC in different regions of the bulge, and found the extinction curve to have at least two degrees of freedom. And \citet{min14} have used VVV aperture photometry complemented with a first version of our new PSF photometry to claim the existence of a Great Dark Lane in front of the Galactic bulge. 

A second, less-prominent overdensity appears about one magnitude below. The nature of this secondary peak is less certain. While \citet{nat11} identify it as the RGB bump, \citet{gon11b} noticed that as the RC becomes brighter at positive longitudes at a given latitude, this secondary clump gets dimmer, raising questions against its identification as the RGB bump. They suggested an alternative explanation as a feature related to the background disk. As our new PSF photometry is more complete, we can better sample this feature and we are working toward its better characterization (Gonzalez et al., in prep.).

Finally, there is also another minor overdensity at brighter magnitudes, at $K_s\sim11.7$ and $J-K_s\sim1.8$ ($Z-K_s\sim3.5$), which also extends toward redder colors and dimmer magnitudes due to the differential extinction. The origin of this overdensity also deserves further exploration. It can be related to the asymptotic giant branch (AGB) bump of the bulge \citep{weg13,nat13}, or to the RGB of the Galactic disk, as we will see in Section \ref{sec_cmdDisk}.

The Besan\c{c}on synthetic CMD for this region (lower right panel in Figure \ref{fig_cmdBulge}) shows the corresponding sequences described in the observational CMDs, but the bulge population seems to be heavily under-represented, with its RGB and RC barely apparent, while the RC of the disk is greatly over-represented.

\subsection{CMDs of the Galactic disk fields in the VVV}
\label{sec_cmdDisk}
The area surveyed in the disk is located at very low Galactic latitudes ($|b|<2\fdg25$). Although our new PSF photometry allows the identification of new low-extinction windows \citep{min18}, these low latitudes imply that most of the CMDs suffer the effect of heavy extinction.
The small range in Galactic latitude and large range in longitude covered (see Figure \ref{fig_dens}) also implies that the most noticeable changes in the CMDs are when we move in Galactic longitude with respect to the Galactic bulge area.

\begin{figure*}
   \centering
   \includegraphics[scale=0.85]{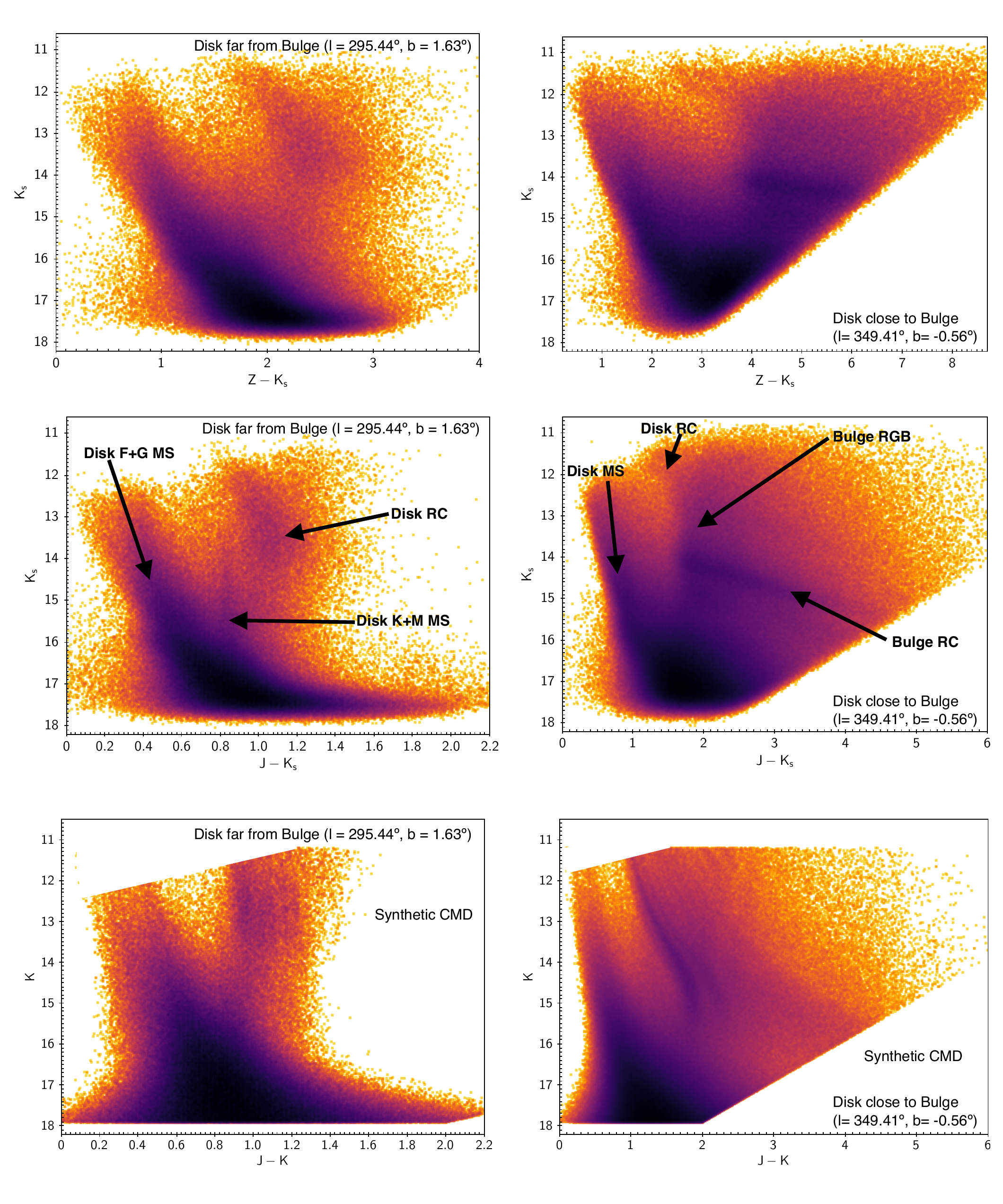}
   \caption{As Figure \ref{fig_cmdBulge}, but in Galactic disk regions for VVV representative fields at the furthest (left panel, at VVV field d115, centered at $l=295\fdg44$, $b=1\fdg63$) and closest (right pane, at VVV field d076, centered at $l=349\fdg41$, $b=-0\fdg56$) longitudes from the Galactic bulge available to the VVV observations. In the CMD in the left panel, we are able to clearly identify the MS and RC of the Galactic disk, even though extinction at low latitudes broaden sequences from further away stars, while in the CMD in the right panel, confusion increases as the Galactic bulge populations become more dominant. A series of online movies show the CMDs for all the different VVV disk fields. }
   \label{fig_cmdDisk}
\end{figure*}

When looking at the CMD located at the lowest VVV longitudes in the disk region, and hence further away from the Galactic bulge (see left panel of Figure \ref{fig_cmdDisk}), we can identify three evolutionary sequences. The most populated one corresponds to the  MS of the disk population, and extends from $K_s\sim12.5$ and $J-K_s\sim0.2$ ($J-K_s\sim0.4$) down to  $K_s\sim18$ and $J-K_s\sim0.9$ ($Z-K_s\sim2$), although for the faintest magnitudes it significantly broadens in color due to the photometric errors. The RGB stars from the disk are located to the red of this sequence. The RC stars are dominant among the RGB. But differently from the RC bulge population, the RC does not present a clumpy structure in the CMD, but a more extended one, stretching from $K_s\sim12$ and $J-K_s\sim0.9$ ($J-K_s\sim1.8$) down to  $K_s\sim14$ and $J-K_s\sim1.1$ ($Z-K_s\sim2.5$). This is mainly due to the stratification at different distances of these stars in the Galactic disk. The RC gets a little redder as it dims, showing that extinction also affects the light of the RC stars located further away from us. RC stars in the disk have proven to be a useful way in tracing its shape. Using the RC stars found with our new PSF photometry in the newly discovered low-extinction Dante's window, \citet{min18} has been able to trace different components of the spiral arms, while \citet{min11a} trace the edge of the Galactic disk following the rapid decline and termination of the RC sequence along different lines of sight on the UKIDSS and VVV fields, using for these last ones a first implementation of our PSF photometry. Finally, the sequence of the K and M dwarfs from the thin disk is clearly distinct at $J-K_s\sim0.85$ and $14<K_s<16$, in a similar position as shown in the left panel of Figure \ref{fig_cmdBulge} in Section \ref{sec_cmdBulge}. At magnitudes $K_s<16$ it becomes blurred by its mixture with the MS. In the $K_s$ vs $Z-K_s$ CMD this sequence does not stand out clearly, probably because it is less clumpy in $Z-K_s$ than in $J-K_s$, as we see in the CMDs of the outer bulge (left panels of Figure \ref{fig_cmdBulge}), and it gets confused with stars from the RGB and the MS. The synthetic CMD based on the Besan\c{c}on Galaxy model (lower left panel of Figure \ref{fig_cmdDisk}) shows a good agreement with the observational CMD for some of the evolutionary branches for example, MS and RC, but fails to clearly show the K and M dwarf branch.

When moving toward higher longitudes, the CMDs look increasingly populated by stars from the bulge. The right panels of Figure \ref{fig_cmdDisk} show the presence of bulge RGB stars that are occupying the same regions of the CMD where the disk RGB and the dwarf sequence are located, making very difficult to disentangle and study these sequences at such longitudes with just this tool. The lower right panel on Figure \ref{fig_cmdDisk} shows that, as it was the case for the most central and reddened low-latitude regions in Figure \ref{fig_cmdBulge}, the Besan\c{c}on Galactic model underrepresents the contribution of RGB stars from the bulge and overrepresents the contribution from RC stars from the disk.    

\subsection{CMDs of the star clusters in VVV}
\label{sec_cmd_clusters}
\begin{figure*}
   \centering
   \includegraphics[scale=0.85]{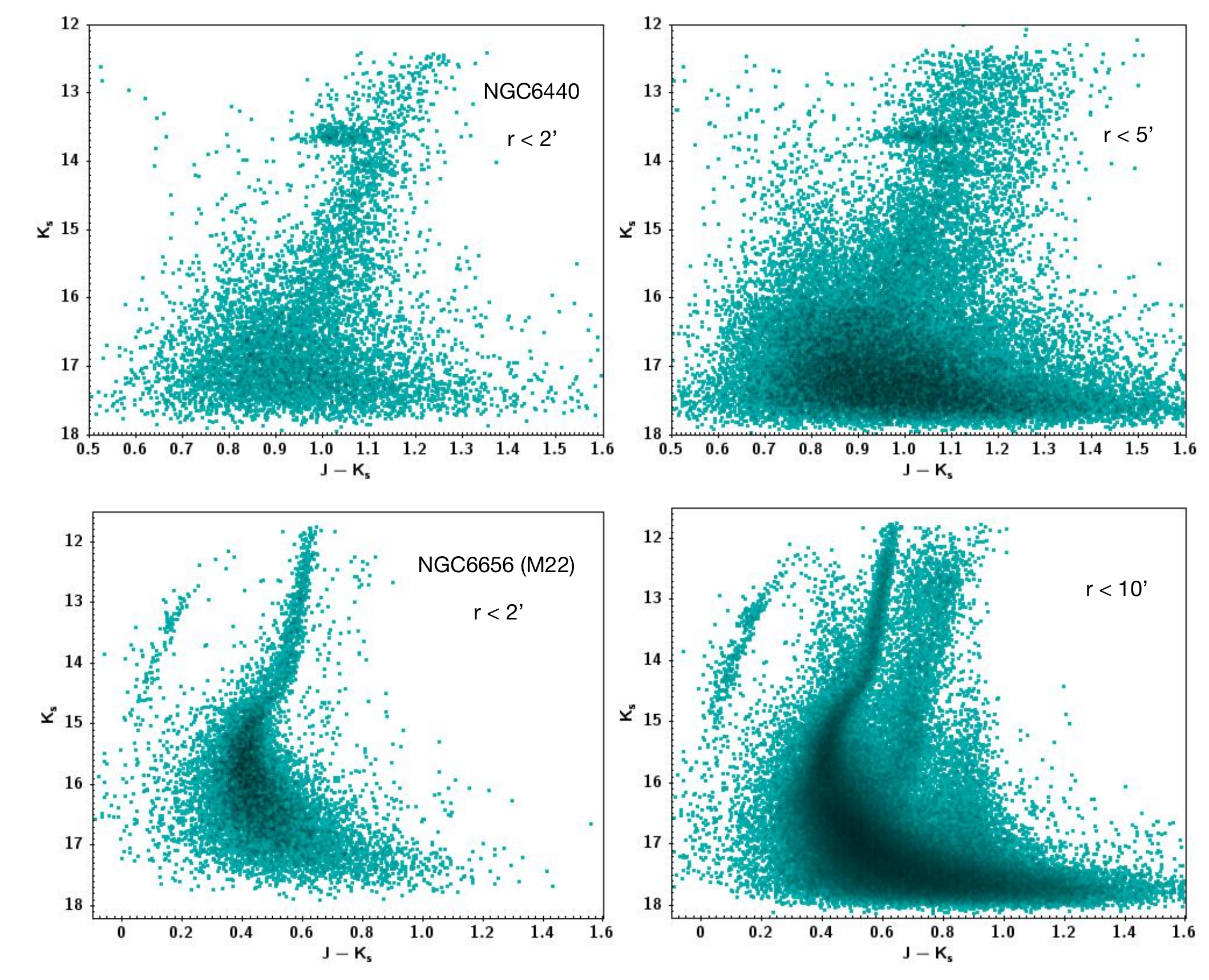}
   \caption{CMDs of two of the GCs surveyed by VVV: in the upper panels, the metal-rich NGC~6440, and in the lower panels, the metal-poor NGC~6656 (M~22). The panels on the left show the inner regions of the GCs were most of the stars present are cluster stars, and the evolutionary sequences of the GC are clearly visible (RGB and RC for NGC~6440; HB, RGB and upper MS for M~22). On the left, we are including stars that are further away from the cluster center. Field stars start to dominate in the CMDs. In the upper right panel, stars from the RGB for NGC~6440 mixed with field bulge stars in the CMD. In the lower panel, the sequences from the bulge are clearly visible too, but the confusion with stars from M~22 is less important since M~22 is much closer than the bulge. }
   \label{fig_cmdclusters}
\end{figure*}

This current release of the PSF photometry allows to build accurate CMDs of the crowded star clusters in the inner region of our Galaxy. There are 36 globular clusters in the most updated version of the Harris catalog \citep{har96} in the area surveyed by VVV. Our photometry allows to build near-infrared CMDs that cover from the centers of the clusters out to their outskirts and immediate surroundings (see Figure \ref{fig_cmdclusters}). Although in only a very few cases are we able to reach magnitudes below the turn-off point (TO) of the MS of these clusters (e.g., M~22, see lower panels of Figure \ref{fig_cmdclusters}), we can certainly sample the brightest sequences, that is, RGB and HB, in a homogeneous way (e.g., NGC6440, see upper panels in Figure \ref{fig_cmdclusters}). Our new photometry also allows for the discovery and characterization of poorly populated globular clusters in the most reddened and crowded regions of our Galaxy \citep{min17c,min17b}. Extending our PSF photometry to all the epochs available for a given cluster allows to inspect and study its population of variable stars, which can supplement additional information to the analysis of the CMDs to derive the physical parameters of these objects \citep{alo15,min17a}. Near-infrared CMDs of the inner Galactic open and young star clusters have also been obtained with this photometry \citep{bor16,nav16,pal16} and they have been used to complement our knowledge of these objects.

\section {Conclusions}
\label{sec_con}
We have presented a new near-infrared photometric catalog of sources detected toward the inner regions of the Milky Way surveyed by VVV, in the $Z$,$Y$,$J$,$H$ and $K_s$ filters. We described the process of extracting the PSF photometry that allowed us to significantly increase the number of detected sources with respect to previous VVV catalogs based on aperture photometry, reaching deeper magnitudes and increasing the completeness factor in the different magnitude ranges. The final catalog contains close to one billion sources, making it the most complete and homogeneous catalog to date of inner Galactic sources. Using this catalog, we built the VVV giga-CMD, a series of near-infrared CMDs depicting the sources contained in the VVV survey, which provide a more complete view of the different stellar populations present in the innermost Milky Way. We qualitatively described the CMDs at different regions of the Galactic bulge and the inner disk, trying to disentangle when possible the stellar populations from the different components of our Galaxy.
We make this photometric database publicly available in the VSA archives. Throughout the article we have highlighted some of the works that have been published using preliminary versions of this database. By sharing the photometry with the whole community, we aim to extend its use and impact on the studies of the inner regions of our Galaxy. This database will be complemented in the near future with the catalogs from the VVVX, the extended VVV survey.

\begin{acknowledgements}
The authors gratefully acknowledge the use of data from the ESO Public Survey program ID 179.B-2002, taken with the VISTA telescope, and data products from the Cambridge Astronomical Survey Unit. This work was supported by the Ministry of Economy, Development, and Tourism's Millennium Science Initiative through grant IC120009, awarded to the Millennium Institute of Astrophysics (MAS), and by he BASAL Center for Astrophysics and Associated Technologies (CATA) through grant PFB-06. J.A-G. also acknowledges support by FONDECYT Iniciaci\'on 11150916 and by the Ministry of Education through grant ANT-1655. R.K.S. acknowledges support from CNPq/Brazil through projects 308968/2016-6 and 421687/2016-9. M.C. acknowledges additional support by FONDECYT grant 1171273. The Geryon cluster housed at the Centro de Astro-Ingenieria UC was used for part of the calculations performed in this paper. The BASAL PFB-06 CATA, Anillo ACT-86, FONDEQUIP AIC-57, and QUIMAL 130008 provided funding for several improvements to the Geryon cluster.
\end{acknowledgements}

%
   \bibliographystyle{aa} 
   \bibliography{mybibtex} 
%




\clearpage

\begin{table*}
  \caption{Effective exposure times of the VVV stacked images used in our analyisis }
  \label{tab_exptimes}
  \centering
  \begin{tabular}{c c c c c c}
    \hline\hline                 
 & $Z$ & $Y$ & $J$ & $H$ & $K_s$ \\
\hline                        
Bulge & 20s & 20s & 24s & 8s & 8s \\
Disk & 40s (epoch1) & 40s (epoch1) & 40s & 40s & 8s \\
 & 20s (epoch2) & 20s (epoch2) & & & \\
   \hline                                   
  \end{tabular}
\end{table*}
    
\end{document}